\pdfoutput=1
\documentclass[prd,preprint,floats,floatfix,aps,nofootinbib,preprintnumbers]{revtex4-1}
\usepackage{graphics,graphicx,psfrag}
\usepackage{amsmath,amssymb}
\usepackage[sort&compress]{natbib}
\usepackage{bm}	
\usepackage{verbatim}
\usepackage{multirow}
\usepackage{subfigure}
\usepackage{ulem}
\usepackage{color}
\usepackage[linktocpage=true,bookmarksnumbered=true,colorlinks=true,plainpages,linkcolor=blue,citecolor=red]{hyperref}
\usepackage{soul}	
\soulregister\cite7
\soulregister\ref7
\soulregister\pageref7
\def\beq{\begin{equation}}
\def\eeq{\end{equation}}
\def\bea{\begin{eqnarray}}
\def\eea{\end{eqnarray}}
\def\bealn{\begin{eqnarray}}
\def\eealn{\end{eqnarray}}
\def\nn{\nonumber}

\def\ifb{\rm fb^{-1}}

\def\tev{\rm TeV}
\def\zp{{Z^\prime}}	
\def\dcol{D_{\rm{col}}}
\def\mres{M_{\rm{res}}}

\def\ie{\textit{i.e.}}
\begin{document}
\preprint{
{\vbox {
\hbox{\bf MSUHEP-140608}
\hbox{\today}
}}}
\vspace*{2cm}

\title{Distinguishing\\Flavor Non-universal Colorons from $Z'$ Bosons\\at the LHC}
\vspace*{0.25in}   
\author{R. Sekhar Chivukula}
\email{sekhar@msu.edu}
\author{Pawin Ittisamai}
\email{ittisama@msu.edu}
\author{Elizabeth H. Simmons}
\email{esimmons@msu.edu}
\affiliation{\vspace*{0.1in}
Department of Physics and Astronomy\\
Michigan State University, East Lansing U.S.A.\\}
\vspace*{0.25 in} 


\begin{abstract}
\vspace{0.5cm}
\noindent
Electrically-neutral massive color-singlet and color-octet vector bosons, which are often predicted in theories beyond the Standard Model, have the potential to be discovered as dijet resonances at the LHC. A color-singlet resonance that has leptophobic couplings needs further investigation to be distinguished from a color-octet one. In previous work, we introduced a method for discriminating between the two kinds of resonances when their couplings are flavor-universal, using measurements of the dijet resonance mass, total decay width and production cross-section.  Here, we describe an extension of that method to cover a more general scenario, in which the vector resonances could have flavor non-universal couplings; essentially, we incorporate measurements of the heavy-flavor decays of the resonance into the method. We present our analysis in a model-independent manner for a dijet resonance with mass $2.5-6.0\,\tev$ at the LHC with $\sqrt{s}=14\,\tev$ and integrated luminosities of  $30,\,100,\,300$ and $1000\,\ifb$, and show that the measurements of the heavy-flavor decays should allow conclusive identification of the vector boson. Note that our method is generally applicable even for a $\zp$ boson with non-Standard invisible decays.  We include an appendix of results for various resonance couplings and masses to illustrate how well each observable must be measured to distinguish colorons from $\zp$s.

\end{abstract}

\maketitle

\section{Introduction}
\label{sec:intro}
Electrically-neutral massive vector resonances are common predictions of models proposed to address issues unsolved by the Standard Model (SM) of particle physics. The resonances that couple to quarks of the SM have the potential to be produced copiously at hadron colliders including the LHC, then decay back into a pair of quarks, yielding a final state of simple topology. The two commonly predicted vector resonances of this kind are a color-octet  and a color-singlet vector boson.

A color-octet vector boson typically arises as a result of extending the gauge group of the strong sector. This means the couplings between quarks and the color-octet can be either flavor universal or flavor non-universal, and chiral or vectorlike. Examples of flavor universal scenarios are the classic axigluon~\cite{Frampton:1987dn, Frampton:1987ut} and coloron~\cite{Chivukula:1996yr, Simmons:1996fz} where all quarks are assigned to be charged under the same $SU(3)$ group in the extended gauge group sector. Flavor non-universal scenarios appear in the case of the topgluon where the third generation quarks are assigned to one $SU(3)$ group and the light quarks to the other~\cite{Hill:1991at, Hill:1994hp}, and the newer axigluon models where different chiralities of the same quark can be charged under different groups~\cite{Antunano:2007da, Ferrario:2009bz, Frampton:2009rk, Rodrigo:2010gm, Chivukula:2010fk, Tavares:2011zg}. Other examples of color-octet vector bosons include Kaluza-Klein (KK) gluons which are excited gluons in extra-dimensional models~\cite{Dicus:2000hm}, technirhos which are composite colored vector mesons found in technicolor~\cite{Farhi:1980xs,Hill:2002ap, Lane:2002sm} models that include colored technifermions, and low-scale string resonances ~\cite{Antoniadis:1990ew}.

An electrically neutral color-singlet vector boson, commonly called a $Z^\prime$, also appears in many scenarios beyond the SM and can originate from extending the electroweak $U(1)$ or $SU(2)$ gauge group. For reviews of $Z^\prime$ models, see Refs.~\cite{Langacker:2008yv, Leike:1998wr, Hewett:1988xc} and the references therein. The $Z^\prime$ can have flavor universal~\cite{Senjanovic:1975rk, Georgi:1989ic, Georgi:1989xz} or flavor non-universal couplings to fermions~\cite{Muller:1996dj, Malkawi:1996fs, Chivukula:1994mn}, where the latter happens when the gauge group for the $\zp$ does not commute with the $SU(2)_L$ of the standard model. A typical $\zp$ can couple to leptons as well as quarks. Those that couple to leptons will decay to charged leptons which have simple and clean experimental signatures, or decay to electrically neutral states, such as the SM's neutrinos, which could register in a detector as missing transverse energy. However, it is possible to have a $\zp$ that does not decay to charged leptons (see, for example,~\cite{Harris:1999ya}) and has to be probed via its hadronic channels such as a dijet final state. In this article, we are interested in $\zp$ bosons of this kind, because one of them could appear as a dijet resonance in a manner similar to a coloron. We denote such a state as a ``leptophobic $\zp$''.

There have been numerous searches for Beyond-the-Standard-Model (BSM) resonances decaying to dijet final states at colliders including the CERN $\mathrm{S\bar{p}pS}$~\cite{Arnison:1986vk, Alitti:1993pn}, the Tevatron~\cite{Abe:1989gz, Abe:1995jz, Abe:1997hm, Abazov:2003tj, Aaltonen:2008dn}, and the Large Hadron Collider (LHC)~\cite{Aad:2010ae, Khachatryan:2010jd, Chatrchyan:2011ns, Aad:2011fq, atlas:2012nma,ATLAS:2012pu,ATLAS:2012qjz,Chatrchyan:2013qha, CMS:kxa}. As no new dijet resonances have been discovered so far, the current exclusion limits on the production cross section for those of sufficiently narrow width have been set by searches carried out by ATLAS and CMS collaborations at the LHC with center-of-mass energy of $8 \,\tev$~\cite{ATLAS:2012pu, Chatrchyan:2013qha, CMS:kxa}. The upgraded, higher-energy LHC will be able to seek a resonance with a larger mass, and the longer projected run time will allow for gathering a sufficient number of signal events to reach the discovery threshold (see, for example,~\cite{Han:2010rf,Yu:2013wta} for recent studies of dijets at the future LHC).

We therefore ask: once a vector resonance has been discovered via the dijet channel, what can we learn about the resonance using information readily available after the discovery?

Many important properties of a resonance are determined by its couplings to the fermions of the Standard Model. In~\cite{Atre:2012gj}, we provided a method for determining the chiral properties of a coloron with flavor-universal couplings to quarks in future LHC runs with center of mass $14\,\tev$ using a combination of the dijet cross section and the rate of coloron production in association with leptonically-decaying standard models weak bosons%
\footnote{The work was inspired by studies of a $\zp$ having leptonic final states~\cite{Cvetic:1992qv, delAguila:1993ym}, which has different decay channels and backgrounds from those of a coloron.}%
. That method is also applicable to determining the chiral couplings of a leptophobic $\zp$. Here, in contrast, we consider how to identify the color structure of a new dijet resonance.

In previous work~\cite{Atre:2013mja}, we introduced a way to distinguish whether a vector resonance is either a leptophobic color-singlet or a color-octet, using a construct that we called a ``color discriminant variable'', $\dcol$. The variable is constructed from the dijet cross-section for the resonance ($\sigma_{jj}$), its mass ($M$), and its total decay width ($\Gamma$), observables that will be available from the dijet channel measurements of the resonance\footnote{Note that $D_{col}$ is dimensionless in the units where $\hbar$ and $c$ equal $1$.}:
	\beq
		\dcol \equiv \frac{M^3}{\Gamma} \sigma_{jj},
	\label{eq:dcol}
	\eeq
For a narrow-width resonance, the color discriminant variable is independent of the resonance's overall coupling strength.  We also illustrated applications of the color discriminant variable technique for two simple cases~\cite{Atre:2013mja}. The first was a flavor universal model with identical couplings to all quarks. In the second, the overall strength of couplings to quarks in the third generation was allowed to be different from those in the first two (couplings to top and bottom were kept equal). Combining $D_{col}$ with information about the $t\bar{t}$ cross-section%
\footnote{Studies of the sensitivity to a high-mass resonance decaying to $t\bar{t}$ at the $\sqrt{s}=14\,$Tev LHC include~\cite{ATL-PHYS-PUB-2013-003, ATLAS:2013hta, CMS:2013xfa, Agashe:2013hma, Iashvili:2013ada}.} %
for the resonance ($pp \to V \to t\bar{t}$) still enabled us to distinguish $C$ from $\zp$ in the second case. While these two scenarios clearly illustrate the application of the method, they did not encompass the features of a typical $\zp$, whose up- and down-type couplings are usually also different from one another.

In this paper we demonstrate the application of the $D_{col}$ method to more general scenarios where couplings to quarks within the same generation (e.g., up vs. down, left-handed vs right-handed) are different, while still allowing couplings to quarks in the third generation to be different from those in the first two. This general scenario corresponds to more realistic models, especially those featuring a $\zp$, as in the case where the gauge group responsible for the existence of the resonance does not commute with the gauge groups of the Standard Model. We discuss how one could incorporate information from heavy resonance decays to a top pair ($t\bar{t}$) and a bottom pair ($b\bar{b}$) in order to determine what type of resonance has been discovered.\footnote{Studies of the $b$-tagging efficiencies of the LHC detectors high-p$_T$ may be found here \cite{ATL-PHYS-PUB-2014-013,ATL-PHYS-PUB-2014-014,CMS:2013vea}.} In particular, we show that in the region of parameter space in which a high-mass coloron or $\zp$ can be discovered as a dijet responsnce at the LHC (with luminosities up to 1000 fb$^{-1}$), a measurement of $\dcol$ at the level of 50\% and a measurement of the $t\bar{t}$ and $b\bar{b}$ cross sections to order one are sufficient to distinguish between color-octet and color-singlet resonances regardless of the details of the flavor structure.


The rest of the paper is organized as follows. We lay out the phenomenological parameters we use, as well as key assumptions imposed on them, in section \ref{sec:genpar}. Then, after briefly reviewing the color discriminant variable in a flavor-universal context in section \ref{subsec:flav-univ}, we discuss the detailed application to flavor-nonuniversal scenarios in section \ref{subsec:flavnon-univ-gen}. The parameter space to which our method is applicable is then presented in section \ref{sec:paramspace}. After reprising, in section \ref{sec:sens}, the estimation of uncertainties for the LHC at $\sqrt{s} = 14\,\tev$ from \cite{Atre:2013mja}, we present our results in section \ref{sec:result-dcols} and discuss them in section \ref{sec:discussions}. A discussion of the uncertainties arising from the parton distribution functions is given in Appendix~\ref{sec:uncertainties_due_to_pdf}.
Model-independent plots for various combinations of resonance couplings and masses are presented in Appendix~\ref{sec:exact_dcol} to illustrate how well each relevant observable must be measured to distinguish between the coloron and the $\zp$.

\section{General Parameterization and Assumptions}
\label{sec:genpar}
In this section, we introduce the parametrization of the couplings of the vector resonances as well as assumptions about properties regarding chiral and generation structures that are used in this article.

We parametrize coloron and leptophobic $\zp$ couplings from a phenomenological point of view. A coloron ($C$) or a $\zp$ that leads to a dijet resonance is produced at hadron colliders via quark-antiquark annihilation%
\footnote{The resonances corresponding to these particles are not produced by gluon fusion: the $\zp$ is not colored and the coloron does not couple to gluon pairs (except very weakly at one loop and higher orders \cite{Chivukula:2013xla}).}%
. The interaction of a $C$ with the SM quarks $q_i$ is described by
	\beq
		\mathcal{L}_C  = i g_{QCD} C_\mu^a \sum_{i=u,d,c,s,t,b}
		\bar{q}_i\gamma^\mu t^a \left( g_{C_L}^i P_L + g_{C_R}^i P_R \right) q_i , \\
	\label{eq:colcoupl}
	\eeq
where $t^a$ is an $ \text{SU}(3) $ generator, $g_{C_L}^i$ and $g_{C_R}^i$ denote left and right chiral coupling strengths, relative to the strong coupling $g_{QCD}$, of the color-octet to the SM quarks. The projection operators have the form $P_{L,R} = (1 \mp \gamma_5)/2$ and the quark flavor index $i$ runs over $ i=u,d,c,s,t,b.$ Similarly, the interactions of a leptophobic $\zp$ with the SM quarks are given by
	\beq
		\mathcal{L}_{\zp}  = i g_w  \zp_\mu \sum_{i=u,d,c,s,t,b}
		\bar{q}_i \gamma^\mu\left( g_{\zp_L}^i P_L + g_{\zp_R}^i P_R \right) q_i,
	\label{eq:zpcoupl}
	\eeq
where $g_{\zp_L}^i$ and $g_{\zp_R}^i$ denote left and right chiral coupling strengths of the leptophobic $\zp$ to the SM quarks relative to the weak coupling $g_w = e/\sin\theta_W$.

Couplings between the vector boson and the left- and right-handed forms of the up- and down-type fermions are not necessarily equal in general. It is true that a color octet resonance, originating from interactions described by a gauge group commuting with $SU(2)_L$ of the standard model, will have the same couplings to the left-handed up- and down-type quarks. The same is not generally  the case for a $\zp$ boson. Moreover, the couplings of either a coloron or $\zp$ to right-handed up- and down-type quarks may differ.

In addition, the couplings can generally be different among the three generations of quarks. The observed suppressions of flavor-changing neutral currents disfavor a TeV-scale resonance with non-universal couplings to the first two generations%
\footnote{
See, for example, Table 4 of Ref.~\cite{Bona:2007vi}.
}. %
So we will limit our interests throughout this article to scenarios where couplings for the first two generations are the same. The third generation is special. In models where top quark plays a unique role, the couplings to quarks in the third generation are often assumed to be different from those for the light quark generations.

Therefore, under this assumption, a coloron has 6 free parameters describing its couplings to quarks:
	\bea
		g^{u,c}_{C_L} = g^{d,s}_{C_L} \qquad&\text{and}&\qquad g^{u,c}_{C_R},\, g^{d,s}_{C_R} 	\nonumber\\
		g^{t}_{C_L} = g^{b}_{C_L} \qquad&\text{and}&\qquad g^{t}_{C_R},\, g^{b}_{C_R}\,,
	\eea
while a leptophobic $Z^\prime$ has 8:
	\bea
		g^{u,c}_{\zp_L} ,\, g^{d,s}_{\zp_L} \qquad&\text{and}&\qquad g^{u,c}_{\zp_R},\, g^{d,s}_{\zp_R} \nonumber\\
		g^{t}_{\zp_L} ,\, g^{b}_{\zp_L} \qquad&\text{and}&\qquad g^{t}_{\zp_R},\, g^{b}_{\zp_R} \,.
	\eea
The dependence on these parameters does not fully manifest itself in measurements available after a discovery; \ie, width and dijet cross section. After all, those observables are not sensitive to chiral structures of the coupling, as the left- and right-handed couplings enter symmetrically. So we denote
	\beq
		g^{q2} \equiv g_{L}^{q2} + g_{R}^{q2}
	\eeq
and notice that the four relevant parameters for our analysis of colorons are
	\beq
		g^{u\,2}_{C} = g^{c \,2}_{C},\quad g^{d\,2}_{C}=g^{s \,2}_{C},\quad g^{t\,2}_{C},\quad g^{b\,2}_{C}
	\eeq
and, similarly, there are four
	\beq
		g^{u\,2}_{\zp} = g^{c \,2}_{\zp},\quad g^{d\,2}_{\zp}=g^{s \,2}_{\zp},\quad g^{t\,2}_{\zp},\quad g^{b\,2}_{\zp}
	\eeq
for a leptophobic $\zp$.

The dijet cross section ($\sigma(pp \to V \to jj)$) plays an important role in evaluating the color discriminant variable. In this analysis, we make a distinction between quarks from the first two ``light'' generations and those from the third. So we will classify what is referred to as ``dijet'' resonance accordingly. Not only does this simplify the analysis, as we shall see later on, but measurements of the cross sections to the $t\bar{t}$ and $b\bar{b}$ final states (respectively, $\sigma(pp \to V \to t\bar{t})$ and $\sigma(pp \to V \to b\bar{b})$) will provide distinct information allowing the identification of the color structure of the resonance. Throughout the article, quarks that constitute a jet $j$ are those from the first two generations; namely,
	\beq
		j = u,\, d,\, c,\, s.
	\eeq
With these definitions in mind, we now discuss how to construct a color discriminant variable.

\section{Defining Color Discriminant Variables in Flavor Non-universal Models}
\label{sec:coldis}
In this section, we briefly review the idea behind the color discriminant variable using a flavor-universal scenario presented in~\cite{Atre:2013mja}. Then we introduce the color discriminant variable for more general scenarios of a resonance with flavor non-universal couplings.

A vector boson coupled to quarks in the standard model is capable of being produced in a great abundance at a hadron collider once it reaches the required energy, appearing as a resonance. Then it decays to a final state of simple topology: a pair of jets, top quarks, or bottom quarks, both of which are highly energetic and clustered in the central region of the detector. Once a sufficient number of events is collected, a resonance with a relatively small width will appear as a distinct bump over a large, but exponentially falling, QCD background. These features make the hadronic decay channels favorable for discovery.

Searches for new particles currently being conducted at the LHC are focused on resonances having a narrow width. So one can expect that if a new dijet resonance is discovered, the  dijet cross section, mass and width of the resonance will be measured. These three observables are exactly what is needed to construct the color discriminant variable, as defined in (\ref{eq:dcol}). This variable is independent of the resonance's overall coupling strength and emphasizes the difference in color structures between a coloron and a $\zp$. The expression for the variable in a flavor-universal scenario (see~\cite{Atre:2013mja}) is particularly simple and will be illustrated next.

\subsection{Review of the flavor universal scenario}
\label{subsec:flav-univ}
We briefly review the idea behind the color discriminant variable using a flavor-universal scenario as an illustration. Throughout this article we will work in the limit of sufficiently small width ($\Gamma/M \ll 1$) such that the dijet cross section for a process involving a vector resonance $V$ can be written using a narrow-width approximation
	\beq
		\sigma_{jj}^V \equiv \sigma(pp \rightarrow V \rightarrow jj) \simeq \sigma(pp \to V) Br(V \to jj),
	\label{eq:genxsec}
	\eeq
where $\sigma(pp \to V)$ is the cross section for producing the resonance. Note that $Br(V \to jj)$ is the boson's dijet branching fraction, which equals $4/6$ for a flavor universal vector resonance that is heavy enough to decay to top quarks. Here, we are interested in multi-TeV resonances because many lighter states have already been experimentally excluded. So, in this limit, the total decay width for a heavy coloron is	\beq
		\Gamma_C = \frac{\alpha_s}{2} M_C g_C^2,
	\label{eq:colwiduniv}
	\eeq
and for a leptophobic $Z^\prime$ is

	\beq
		\Gamma_\zp = 3 \alpha_w M_\zp g_\zp^2\,,
	\label{eq:zpwiduniv}
	\eeq
where $g_{C/\zp}^2 = \left( g_{{C/\zp}_L}^2 + g_{{C/\zp}_R}^2\right)$ denotes the flavor-universal coupling of the resonance to quarks. These, respectively, lead to the dijet cross section for a coloron
	\bea
	\nn
		\sigma_{jj}^C &=& \frac{4}{9}  \alpha_s g_C^2 \frac{1}{M_C^2} 	\sum_{q} W_q(M_c)  Br(C \to jj)\\
		&=& 	\frac{8}{9} \frac{\Gamma_C}{M_C^3} \sum_{q} W_q(M_C)  Br(C \to jj),
	\label{eq:colxsec}
	\eea
and for a leptophobic $\zp$,
	\bea
	\nn
		\sigma_{jj}^\zp &=& \frac{1}{3}  \alpha_w g_\zp^2 \frac{1}{M_{\zp}^2} \sum_{q} W_q(M_\zp)  Br(\zp \to jj)\\
	&=& 	\frac{1}{9} \frac{\Gamma_{\zp}}{M_{\zp}^3} \sum_{q} W_q(M_\zp)  Br(\zp \to jj)\,.
	\label{eq:zpxsec}
	\eea
Here the function $W_q$, which is constructed from the parton luminosity for the production of the vector resonance with mass $M_V$ via $q\bar{q}$ annihilation at the center-of-mass energy squared $s$, is defined by
	\beq
		W_{q}(M_V) = 2\pi^2 \frac{M_V^2}{s} \int_{M_V^2/s}^{1} \frac{dx}{x}
			\left[ f_q\left(x, \mu_F^2\right) f_{\bar{q}}\left( \frac{M_V^2}{sx}, \mu_F^2 \right) +
			f_{\bar{q}}\left(x, \mu_F^2\right) f_q\left( \frac{M_V^2}{sx}, \mu_F^2 \right) \right]  \,,
	\label{eq:w_function}
	\eeq
where $f_{q}\left(x,\mu_F^2\right)$ is the parton distribution function at the factorization scale $\mu_F^2$. Throughout this article, we set $\mu_F^2 = M_V^2$.

The fact that the overall coupling strength can be factored out as a ratio of observables ($\Gamma_V/M_V$) as shown in (\ref{eq:colwiduniv}) and (\ref{eq:zpwiduniv}) motivates the definition of the color discriminant variables, which are
	\bea
		\dcol^C &=&  \frac{M_C^3}{\Gamma_C} \sigma_{jj}^C= \frac{8}{9} \left[\sum_{q} W_q(M_C)  Br(C \to jj) \right]
	\label{eq:coldcol}\\
		\dcol^\zp &=& \frac{M_\zp^3}{\Gamma_\zp} \sigma_{jj}^\zp= \frac{1}{9} \left[\sum_{q} W_q(M_\zp)  Br(\zp \to jj)\right]
	\label{eq:zpdcol}
	\eea
for the coloron and $\zp$, respectively.

The factors in the square brackets in (\ref{eq:coldcol}) and (\ref{eq:zpdcol}) are the same for flavor-universal resonances having a particular mass; only the initial numerical factors differ. In other words, the difference between the values of color discriminant variables corresponding to the two types of flavor-universal resonance
	\beq
		\dcol^C = 8 \dcol^\zp
	\label{eq:factor8_dcol}
	\eeq
will help pinpoint the nature of the color structure of the discovered particle. Turning this argument around, a set of measurements of yielding a particular value of $\dcol$ will correspond to a $\zp$ that is $8$ times broader than a coloron of the same mass that is produced at the same rate; \textit{i.e.},
	\beq
		\Gamma_\zp^\star = 8 \Gamma_C^\star \,,
	\eeq
where the star ($^\star$) denotes that particular width.

In a flavor non-universal scenario, one cannot always factor out the dependence of couplings appearing in a production cross section in the manner displayed in equations (\ref{eq:colxsec}) and (\ref{eq:zpxsec}). Branching fractions of the decay final states are also not necessarily the same for two resonances. In the following sections, we will demonstrate that even when this is the case, the color discriminant variable method remains valuable.

\subsection{Flavor non-universal scenario}
\label{subsec:flavnon-univ-gen}
In a flavor non-universal scenario, we will follow the parameterization and assumptions introduced in section \ref{sec:genpar}.  The production cross section and decay width for the coloron are
	\bea
		\sigma(pp\rightarrow C) &=& \frac{4}{9}\frac{\alpha_s}{M_C^2}
			\left[
			 g_{C}^{u\,2} \left( W_u + W_c\right)
			+
			 g_{C}^{d\,2} \left( W_d + W_s\right)
			+
			g_{C}^{b\,2}  W_b
			\right]
		\nonumber
		\\
		 &=& \frac{4}{9}\frac{\alpha_s}{M_C^2}
			\left(  g_{C}^{u\,2} +  g_{C}^{d\,2} \right)
			 \Bigg[
			\frac{g_{C}^{u\,2}}{g_{C}^{u\,2} +  g_{C}^{d\,2} } \left( W_u + W_c\right)
			\nonumber\\
			&& \qquad
			+ \left( 1 - \frac{g_{C}^{u\,2}}{g_{C}^{u\,2} +  g_{C}^{d\,2} } \right) \left( W_d + W_s\right)
			+
			\frac{g_{C}^{b\,2}}{g_{C}^{u\,2} +  g_{C}^{d\,2} } W_b
			\Bigg] \,, \label{eq:c_production}
		\\
		\Gamma_C &=& \frac{\alpha_s}{12}M_C \left[
			2 g_{C}^{u\,2} + 2g_{C}^{d\,2}
			+ g_{C}^{t\,2}+ g_{C}^{b\,2}
			 \right]
		\nonumber\\
		&=& \frac{\alpha_s}{12}M_C \left(g_{C}^{u\,2} +  g_{C}^{d\,2} \right)
			\left[
			2+
			\frac{g_{C}^{t\,2}}{ g_{C}^{u\,2} +  g_{C}^{d\,2} }+ \frac{g_{C}^{b\,2}}{g_{C}^{u\,2} +  g_{C}^{d\,2} }
			 \right] \,,
	\eea
where they have been written using a parametrization that allows some of them to correspond to observables, as we shall see shortly. The expressions for leptophobic $\zp$ are similar
	\bea
		\sigma(pp\rightarrow \zp) &=& \frac{1}{3}\frac{\alpha_w}{M_\zp^2}
			\left(g_{\zp}^{u\,2} + g_{\zp}^{d\,2} \right)
			\Bigg[
			 \frac{g_{\zp}^{u\,2}}{g_{\zp}^{u\,2} + g_{\zp}^{d\,2} } \left( W_u + W_c\right)
			\nonumber\\
			&&	\qquad
			+
			\left( 1 - \frac{g_{\zp}^{u\,2}}{ g_{\zp}^{u\,2} + g_{\zp}^{d\,2} } \right) \left( W_d + W_s\right)
			+
			\frac{g_{\zp}^{b\,2}}{ g_{\zp}^{u\,2} + g_{\zp}^{d\,2} } W_b
			\Bigg] \,, \label{eq:zp_production}
		\\
		\Gamma_\zp &=& \frac{\alpha_w}{2}M_\zp \left(g_{\zp}^{u\,2} + g_{\zp}^{d\,2} \right)
			 \left[
			2
			+ \frac{g_{\zp}^{t\, 2}} {g_{\zp}^{u\,2} + g_{\zp}^{d\,2}}  + \frac{g_{\zp}^{b\, 2}}{g_{\zp}^{u\,2} + g_{\zp}^{d\,2}}
			 \right] \,.
	\eea
The color discriminant variables are, for the coloron,
	\bea
		\dcol^C &=&  \frac{16}{3} \left( W_u + W_c\right)
		\Bigg[
				 \frac{g_{C}^{u\,2}}{g_{C}^{u\,2} +  g_{C}^{d\,2} }
				+
				\left( 1 - \frac{g_{C}^{u\,2}}{g_{C}^{u\,2} +  g_{C}^{d\,2} } \right)  \left(\frac{ W_d + W_s }{ W_u + W_c }\right)
			\nonumber
			\\
			&&\qquad	+
				\frac{g_{C}^{b\,2}}{g_{C}^{u\,2} +  g_{C}^{d\,2} }  \left(\frac{W_b }{ W_u + W_c }\right)
		\Bigg]
		\times
		\left\{
		\frac{
			2
			}
			{
				\left( 2 +
				\frac{g_{C}^{t\,2}}{ g_{C}^{u\,2} +  g_{C}^{d\,2} }+ \frac{g_{C}^{b\,2}}{g_{C}^{u\,2} +  g_{C}^{d\,2} }
				\right)^2
			}
		\right\}
	\label{eq:dcolcol-nonu}
	\eea
and for the $\zp$,
\bea
		\dcol^\zp &=&  \frac{2}{3} \left( W_u + W_c\right)
		\Bigg[
				  \frac{g_{\zp}^{u\,2}}{g_{\zp}^{u\,2} + g_{\zp}^{d\,2} }
				+
				\left( 1 -  \frac{g_{\zp}^{u\,2}}{g_{\zp}^{u\,2} + g_{\zp}^{d\,2} } \right)  \left(\frac{ W_d + W_s }{ W_u + W_c }\right)
			\nonumber
			\\
			&&\qquad	+
				\frac{g_{\zp}^{b\,2}}{g_{\zp}^{u\,2} + g_{\zp}^{d\,2} }   \left(\frac{W_b }{ W_u + W_c }\right)
		 \Bigg]
			\times
			\left\{
			\frac{
			2
			}
			{
				\left(
				2
			+ \frac{g_{\zp}^{t\, 2}} {g_{\zp}^{u\,2} + g_{\zp}^{d\,2}}  + \frac{g_{\zp}^{b\, 2}}{g_{\zp}^{u\,2} + g_{\zp}^{d\,2}}
				\right)^2
			}
			\right\}
	\label{eq:dcolzp-nonu}
	\eea
where parts related to resonance production are grouped within the square brackets, while those related to decay are grouped within curly braces. Notice that the appearance of the factor $2$ in the decay part of the expressions is due to our assumption that the first two generations couple identically to the vector resonance.

The relative strength with which the vector boson couples to the $u-$ and $d-$type quarks of the light SM generations,  $g_u^{2} / \left( g_u^{2} + g_d^{2} \right)$, which we will call the ``up ratio'' for brevity, is not accessible by experiments available in the dijet channel. However, equivalent information for quarks in the third generation can be measured by comparing the dijet and heavy flavor cross sections. Defining cross sections for $t\bar{t}$ and $b\bar{b}$ final states for heavy boson decay
	\bea
	\sigma_{t\bar{t}}^V \equiv \sigma(pp\rightarrow V \rightarrow t\bar{t})
		= \sigma(pp\rightarrow V) \frac{
			g_{t}^{2} / \left( g_{u}^{2} + g_{d}^{2} \right)
		} {
			2+
			\frac{g_{C}^{t\,2}}{ g_{C}^{u\,2} +  g_{C}^{d\,2} }+ \frac{g_{C}^{b\,2}}{g_{C}^{u\,2} +  g_{C}^{d\,2} }
		}
	\eea
and
	\bea
	\sigma_{b\bar{b}}^V \equiv \sigma(pp\rightarrow V \rightarrow b\bar{b})
		= \sigma(pp\rightarrow V) \frac{
			g_{b}^{2} / \left( g_{u}^{2} + g_{d}^{2} \right)
		} {
			2+
			\frac{g_{C}^{t\,2}}{ g_{C}^{u\,2} +  g_{C}^{d\,2} }+ \frac{g_{C}^{b\,2}}{g_{C}^{u\,2} +  g_{C}^{d\,2} }
		}
	\eea
where $\sigma(pp\rightarrow V)$ ($V=C,\,\zp$) has been defined in Eq.~(\ref{eq:c_production}) and (\ref{eq:zp_production}), we find%
\footnote{Hereafter, the superscript $V$ on $\sigma_{}^V$ and the subscript $V$ on $M_V$ and $\Gamma_V$ will be omitted when the meaning is clear from the context.}%
	\bea
		 \frac{g_{t}^{2}} {g_{u}^{2} + g_{d}^{2}} = 2 \frac{\sigma_{t\bar{t}}^V}{\sigma_{jj}^V}\qquad {\rm ``top\ ratio''}
		 \label{eq:top_ratio}
	\eea
and
	\bea
		 \frac{g_{b}^{2}} {g_{u}^{2} + g_{d}^{2}} = 2 \frac{\sigma_{b\bar{b}}^V}{\sigma_{jj}^V} \,, \qquad {\rm ``bottom\ ratio''} \,.
	\eea
Supplementary measurements of these ratios of cross sections will help pinpoint the structure of couplings of the resonance.

\begin{figure}[ht]
{
\includegraphics[width=0.50\textwidth, clip=true]{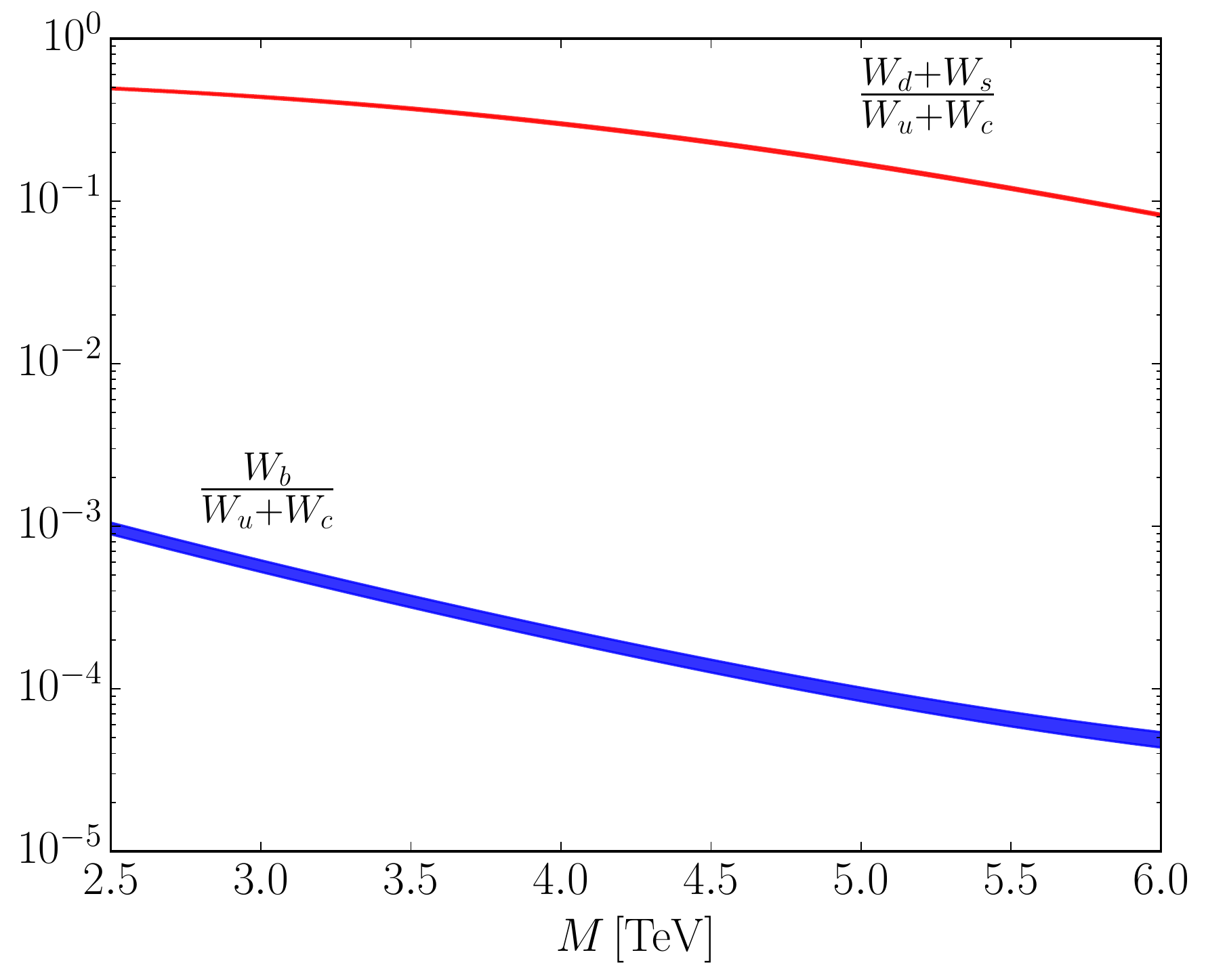}
}
\caption{Ratios of parton density functions as a function of the mass of the produced vector resonance.  This figures shows two such ratios: the relative contribution of parton density functions, defined in Eq.~(\ref{eq:w_function}), for the down-type quarks of the first two generations $\left(\frac{W_d + W_s }{ W_u + W_c }\right)$ (top curve, in red) and for bottom quarks $ \left(\frac{ W_b }{ W_u + W_c }\right)$ (bottom curve, in blue) relative to the up-type quarks for the first two generations. The values have been calculated using CT09MCS parton distribution functions with factorization scale varied by a factor of $2$ away from the mass of the resonance in the range $2.5-6.0\,\tev$. The results of this variation are illustrated as a band for each function.  Note that the upper curve depends only weakly on the resonance mass and that the lower curve's values are
$\mathcal{O}(10^{-3})$ or less over the entire mass range.
}
\label{fig:w-ratio}
\end{figure}

While our expressions (\ref{eq:dcolcol-nonu}, \ref{eq:dcolzp-nonu}) for $\dcol$ appear to have a complicated dependence on multiple parameters involving different quark flavors, recalling that the coloron and Z' are being produced by collisions of quarks lying inside protons simplifies matters considerably.  First, the contribution of b-quarks to the production part of $\dcol$ [square brackets within (\ref{eq:dcolcol-nonu}) and (\ref{eq:dcolzp-nonu})] is suppressed significantly by their relative scarcity in the protons, as reflected by the range of the ratios of the parton density functions $\frac{ W_d + W_s }{ W_u + W_c }$ and $\frac{ W_b }{ W_u + W_c }$. We plot the values of these functions in Fig.~\ref{fig:w-ratio} for mass range of $2.5 - 6\,\tev$ at a $pp$ collider with center-of-mass energy $14\,\tev$.
Here and throughout the paper, we use the CT09MCS parton distribution functions\footnote{
We have checked, using CT10NLO PDF \cite{Lai:2010vv}, that the contribution from the resonance production via $b\bar{b}$ annihilation is still smaller by at least $2$ orders of magnitude (compared to production via $u\bar{u}$ and $d\bar{d}$) after the uncertainties on $\frac{ W_d + W_s }{ W_u + W_c }$ and $\frac{ W_b }{ W_u + W_c }$ due to PDF uncertainties are taken into account.}
 from the CTEQ Collaboration~\cite{Lai:2009ne}. In that figure, we allow the factorization scale to vary by a factor of $2$ from the mass of the resonance. The effect of this variation has been illustrated as the width of each band in the plot. From the plot we conclude that unless the resonance has a {\it much} stronger coupling to the $b$ than to quarks in first two generations, the precise strength of the couplings to third-generation quarks becomes relevant to $\dcol$ only through the decay part of the expressions (\ref{eq:dcolcol-nonu}) and  (\ref{eq:dcolzp-nonu}), the part in curly braces.

Second, we see that the experimentally inaccessible parameter that we call the up ratio appears unlikely to leave us confused as to whether a new dijet resonance is a coloron or a leptophobic $\zp$. This can be seen as follows. First, note that in a flavor-universal scenario, there is a factor of 8 difference between $\dcol$ for a coloron and a $\zp$ of identical mass and production cross-section -- see Eq.~(\ref{eq:factor8_dcol}). Therefore, confusion would only arise between models where a very small up-ratio for the coloron suppressed $\dcol$ and those where an up-ratio near the maximum value of 1 for the $\zp$ enhanced $\dcol$.
Consider two resonances having the same top and bottom ratios. A coloron with negligible up ratio will have
\begin{equation}
	\dcol^C \propto \frac{16}{3} \left[
				 \frac{g_{C}^{u\,2}}{g_{C}^{u\,2} +  g_{C}^{d\,2} }
				+
				\left( 1 - \frac{g_{C}^{u\,2}}{g_{C}^{u\,2} +  g_{C}^{d\,2} } \right)  \left(\frac{ W_d + W_s }{ W_u + W_c }\right)
		\right]
		\rightarrow
		\frac{16}{3} \left[
			\frac{ W_d + W_s }{ W_u + W_c }
		\right]
\end{equation}
while a $\zp$ with the up ratio close to $1$ will have
\begin{equation}
 	\dcol^\zp \propto \frac{2}{3}
		\left[
				  \frac{g_{\zp}^{u\,2}}{g_{\zp}^{u\,2} + g_{\zp}^{d\,2} }
				+
				\left( 1 -  \frac{g_{\zp}^{u\,2}}{g_{\zp}^{u\,2} + g_{\zp}^{d\,2} } \right)  \left(\frac{ W_d + W_s }{ W_u + W_c }\right)
		\right]
		\rightarrow
		\frac{2}{3} \,.
\end{equation}
In this case, Fig.~\ref{fig:w-ratio} illustrates that the only place where the ratio $\frac{ W_d + W_s }{ W_u + W_c }$ is small enough to cause confusion is at high masses.
We will illustrate this further in Section \ref{sec:result-dcols}.

Now that we know the parameters that affect the determination of $\dcol$, the general prescription of the analysis goes as follows. After a resonance has been discovered, one uses the measurements of three observables; dijet cross section, mass, and total decay width to evaluate $\dcol$. This particular value of $\dcol$ could correspond to various configurations of flavor-nonuniversal couplings denoted by three coupling ratios; namely, the up ratio, top ratio, and bottom ratio.  The up ratio cannot be experimentally measured, but the top ratio and bottom ratio are accessible by  measuring the $t\bar{t}$ and $b\bar{b}$ cross sections. In many circumstances, these two cross section measurements together with $\dcol$ suffice to identify the color structure of a resonance of given mass and dijet cross section. The illustration of this method and its limitation are presented in Section \ref{sec:result-dcols}.

First, however, we must determine the region of parameter space where the color discriminant variable is relevant: the region where one can discover the resonance and measure $M$, $\Gamma$ and $\sigma_{jj}$ precisely.  This is the topic of the next section.

\section{Accessible Dijet Resonances at the 14 TeV LHC}
\label{sec:paramspace}
We have argued that the color discriminant variable allows one to distinguish whether a resonance decaying to dijets is a coloron or a leptophobic $\zp$ in a model-independent manner; i.e., without analyzing each set of couplings separately. In this section, we describe the region of parameter space to which the method is applicable.\footnote{A detailed discussion was presented in~\cite{Atre:2013mja}.} In this region the resonance has not already been excluded by the current searches, is within the reach of future searches, and has a total width that is measurable and consistent with the designation ``narrow''.

One may deduce the current exclusion limits on resonances in the dijet channel using the limits on the production cross section times branching ratio ($\sigma \times Br(jj)$) from the (null) searches for narrow-width resonances carried out by the ATLAS and CMS collaborations~\cite{ATLAS:2012pu, Chatrchyan:2013qha, CMS:kxa} at $\sqrt{s} = 8\,\tev$. We use the most stringent constraint, which comes from CMS~\cite{CMS:kxa}. As the exclusion limit is provided in the form of $\sigma\times Br(jj) \times (\mathrm{Acceptance})$, we estimate the acceptance of the detector for each value of the resonance mass by comparing, within the same theoretical model, $\sigma \times Br(jj)$ that we calculated vs. $\sigma \times Br(jj) \times (\mathrm{Acceptance})$ provided by CMS. The acceptance is a characteristic of properties of the detector and kinematics, the latter being the same for coloron and $\zp$ to leading order; thus we use throughout our analysis the acceptance deduced from such a comparison  made within a sequential $\zp$ model. The excluded region of parameter space is displayed in gray in Figs. \ref{fig:param_space_col} and \ref{fig:param_space_zp}.

Sensitivity to a dijet resonance in future LHC experiments with $\sqrt{s}=14\,\tev$ depends on the knowledge of QCD backgrounds, the measurements of dijet mass distributions, and statistical and systematic uncertainties. CMS
~\cite{Gumus:2006mxa} has estimated the limits on $\sigma \times Br(jj) \times (\mathrm{Acceptance})$ that will be required in order to attain a $5\sigma$ discovery at CMS with integrated luminosities up to $10\,\ifb$, including both statistical and systematic uncertainties. We obtain the acceptance for CMS at $\sqrt{s} = 14\,\tev$ in the same manner as described in the previous paragraph. The sensitivity for the dijet discovery from $10\,\ifb$ is then scaled to the integrated luminosities $\mathcal{L} = 30,\,100,\,300,\,1000\,\ifb$ considered in our studies (assuming that the systematic uncertainty scales with the squared root of integrated luminosity). The predicted discovery reaches for these luminosities are shown in varying shades of blues for coloron and greens for $\zp$ in Figs. \ref{fig:param_space_col} and \ref{fig:param_space_zp}.

The total decay width also constrains the absolute values of the coupling constants. On the one hand, experimental searches are designed for narrow-width dijet resonances; hence their exclusion limits are not applicable when the resonance is too broad, which translates to about $\Gamma/M = 0.15$ as the upper limit~\cite{Bai:2011ed, Haisch:2011up, Harris:2011bh}. On the other hand, the appearance of (intrinsic) total decay width in the expression for the color discriminant variable requires that the width be accurately measurable; width values smaller than the experimental dijet mass resolution, $M_\mathrm{res}$, cannot be distinguished. The region of parameter space that meets both constraints and is relevant to our analysis is shown in Figs. \ref{fig:param_space_col} and \ref{fig:param_space_zp} as the region between the two dashed horizontal curves labeled $\Gamma \ge 0.15 M$ and $\Gamma \le M_\mathrm{res}$. Regions where the width is too broad or too narrow are shown with a cloudy overlay to indicate that they are not accessible via our analysis.


\subsection{Uncertainties in the measurement of $\dcol$ at the $14\,\tev$ LHC}
\label{sec:sens}

Statistical and systematic uncertainties on dijet cross section, mass, and intrinsic width of the resonance, as well as the uncertainities in the corresponding $t\bar{t}$ and $b\bar{b}$ branching ratios for such a resonsnce,
 will play a key role in determining how well $\dcol$ can discriminate between models at the LHC with  $\sqrt{s} = 14\,\tev$. The actual values of the systematic uncertainties at the LHC with $\sqrt{s} = 14\,\tev$ will be obtained only after the experiment has begun. In this section, we discuss the estimates of the uncertainties that we use in our calculations.

The effect of systematic uncertainties in the jet energy scale, jet energy resolution, radiation and low mass resonance tail and luminosity on the dijet cross section at the 14 TeV LHC was estimated in Ref.~\cite{Gumus:2006mxa}.  It is presented there as a fractional uncertainty (as a function of the mass) normalized to the dijet cross section required to obtain a 5$\sigma$ discovery above background fluctuations. The dijet mass resolution, the uncertainty of the dijet mass resolution, and the uncertainty of the mass itself (due to uncertainty in the jet energy scale), also affect the determination of both the mass and intrinsic width. Table \ref{tab:uncert} lists these estimated uncertainties for $\sqrt{s} = 14$ TeV together with the values from the actual CMS and ATLAS experiments at $\sqrt{s} = 8$ TeV. Here, we use the reported systematic uncertainties from actual LHC data where available and estimate that any future LHC run will be able to
reach at least that level of precision. This estimate is likely to be conservative, since experiments tend to reduce their systematic uncertainties, and improve the precision with which they understand these uncertainties and their efficiencies by using real data.  The values we use are marked in Table \ref{tab:uncert} with asterisks.
In Figs.~\ref{fig:param_space_col} and \ref{fig:param_space_zp},
we also show the contours in the region of parameter space along which the uncertainties listed above in measuring $\dcol$ are $20\%$ and $50\%$ at the LHC with $\sqrt{s} = 14\,\tev$ for the integrated luminosity of $1000\,\ifb$.

\begin{table}[!ht]
\begin{tabular}{| c | c |  c  | c | r |}
\hline
\hline
	Systematic Uncertainty		&
	Value		&
	Mass Range	&
	$\sqrt{s}$	&
	Experiment\\
\hline
	dijet cross section &
	\multirow{2}{*}{$0.28-0.41^\ast$} &
	\multirow{2}{*}{$2.5-6\,\tev$}	&
	\multirow{2}{*}{$14\,\tev$}		&
	\multirow{2}{*}{LHC~\cite{Gumus:2006mxa} \hspace{-0.14cm}}\\
	 uncertainty (fractional) &
	 	&
	 	&
		&
	\\
\hline
	\multirow{3}{*}{Mass resolution}	&
	$0.045 - 0.035^\ast$ &
	 $2.5-6\,\tev$	&
	$8\,\tev$		&
	CMS~\cite{CMS:kxa}
\\
	{}	&
	$0.045  - 0.031$ &
	 $2.5-6\,\tev$	&
	$8\,\tev$		&
	ATLAS~\cite{ATLAS:2012qjz}
\\
		&
	$0.071 - 0.062$ &
	 $2.5-6\,\tev$	&
	$14\,\tev$		&
	LHC~\cite{Gumus:2006mxa}
\\
\hline
	Mass resolution 	&
	$0.1^\ast$ &
	any			&
	$8\,\tev$		&
	CMS~\cite{Chatrchyan:2013qha}
\\
	uncertainty	&
	$0.1$ &
	any			&
	$14\,\tev$		&
	LHC~\cite{Gumus:2006mxa}
\\
\hline
	Mass uncertainty 	&
	$0.013^\ast$ &
	any	&
	$8\,\tev$		&
	CMS~\cite{Chatrchyan:2013qha}
\\
	from jet energy	&
	$0.028$ &
	any	&
	$8\,\tev$		&
	ATLAS~\cite{ATLAS:2012qjz}
\\
	scale	&
	$ 0.035$ &
	any	&
	$14\,\tev$		&
	LHC~\cite{Gumus:2006mxa}
\\
\hline
\hline
\end{tabular}
\caption{Sources of systematic uncertainty contributing to uncertainties in measurement of the cross section, mass and width of a resonance in the dijet channel at various experiments and center of mass energies. These determine how well $\dcol$ can discriminate between models at the LHC with $\sqrt{s} = 14\,\tev$. In this analysis, we use the estimate for systematic uncertainties from actual LHC data where available and assume that any future LHC run will be able to reach at least the current level of precision. The values we used are indicated by an asterisk. Reproduced from~\cite{Atre:2013mja}.}
\label{tab:uncert}
\end{table}

We estimate uncertainty on $\dcol$ due to PDF uncertainties by using the CT10NLO PDF set from the CTEQ collaboration \cite{Lai:2010vv} as described in Appendix~\ref{sec:uncertainties_due_to_pdf}. In a typical scenario where couplings to the up-type quarks are not very small, the uncertainty will range from $\mathcal{O}(5\%)$ for low masses to $\mathcal{O}(30\,\%)$ for high masses. Rather than incorporating PDF uncertainties into our analysis, we instead assess the question of how well $\dcol$ has to be measured to distinguish a coloron from a $\zp$. We consider scenarios where $\dcol$ is measured to within $20\,\%$ and $50\,\%$, the range that should encompass the expected magnigudes of the uncertainties from PDF and other sources mentioned earlier.

While a detailed analysis of the uncdertainties in $\sigma_{t\bar{t}}$ and $\sigma_{b\bar{b}}$ lies beyond the scope of this paper,\footnote{For recent studies of these topics, however, see footnotes 3 and 4.}  we will find that measurements of these cross sections to order one are sufficient for the purpose of discriminating between a coloron and a $\zp$ over much of the interesting parameter space.

\begin{figure}[t]
{
\includegraphics[width=0.31\textwidth, clip=true]{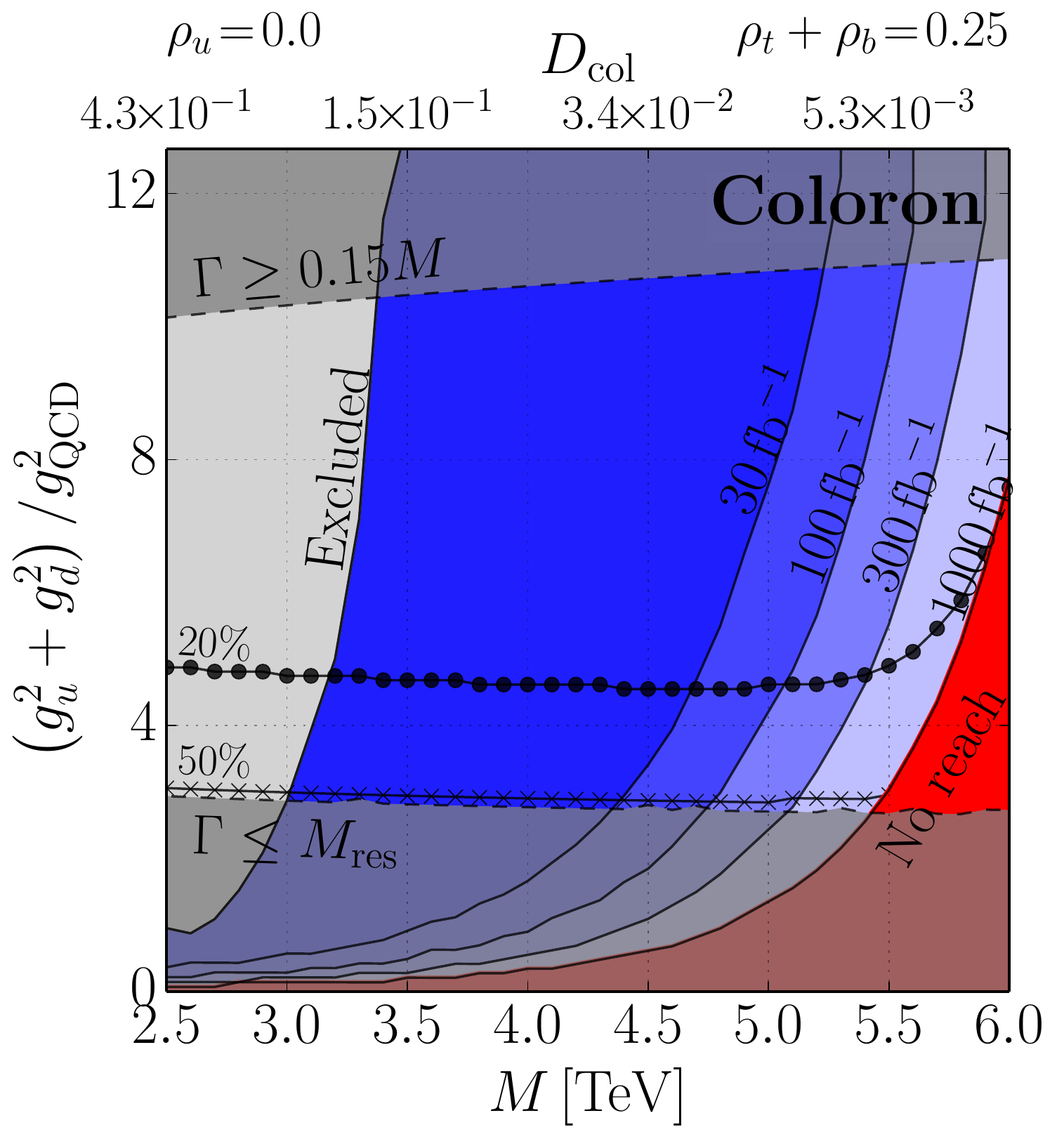}
\includegraphics[width=0.31\textwidth, clip=true]{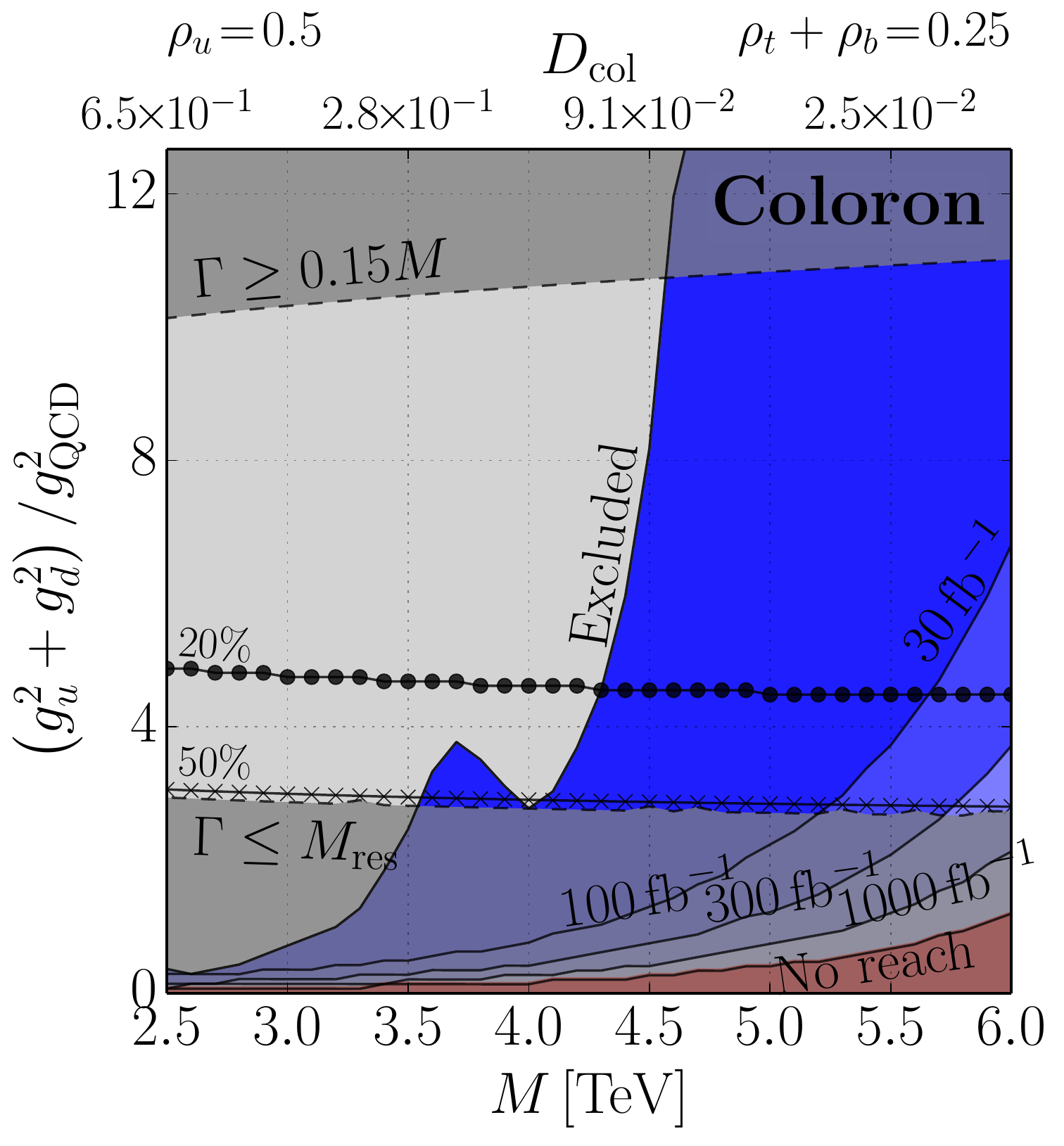}
\includegraphics[width=0.31\textwidth, clip=true]{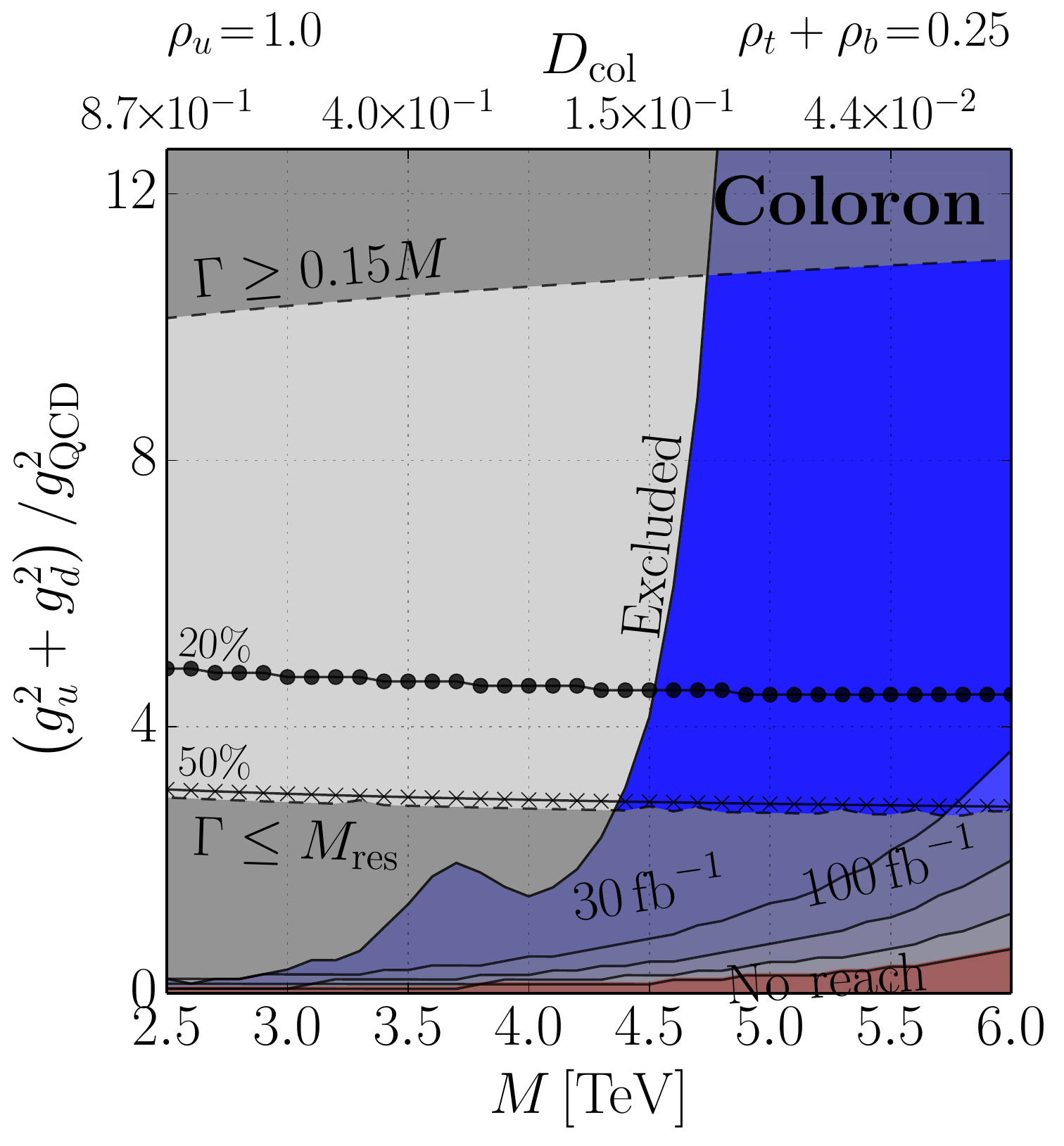}
\\
\includegraphics[width=0.31\textwidth, clip=true]{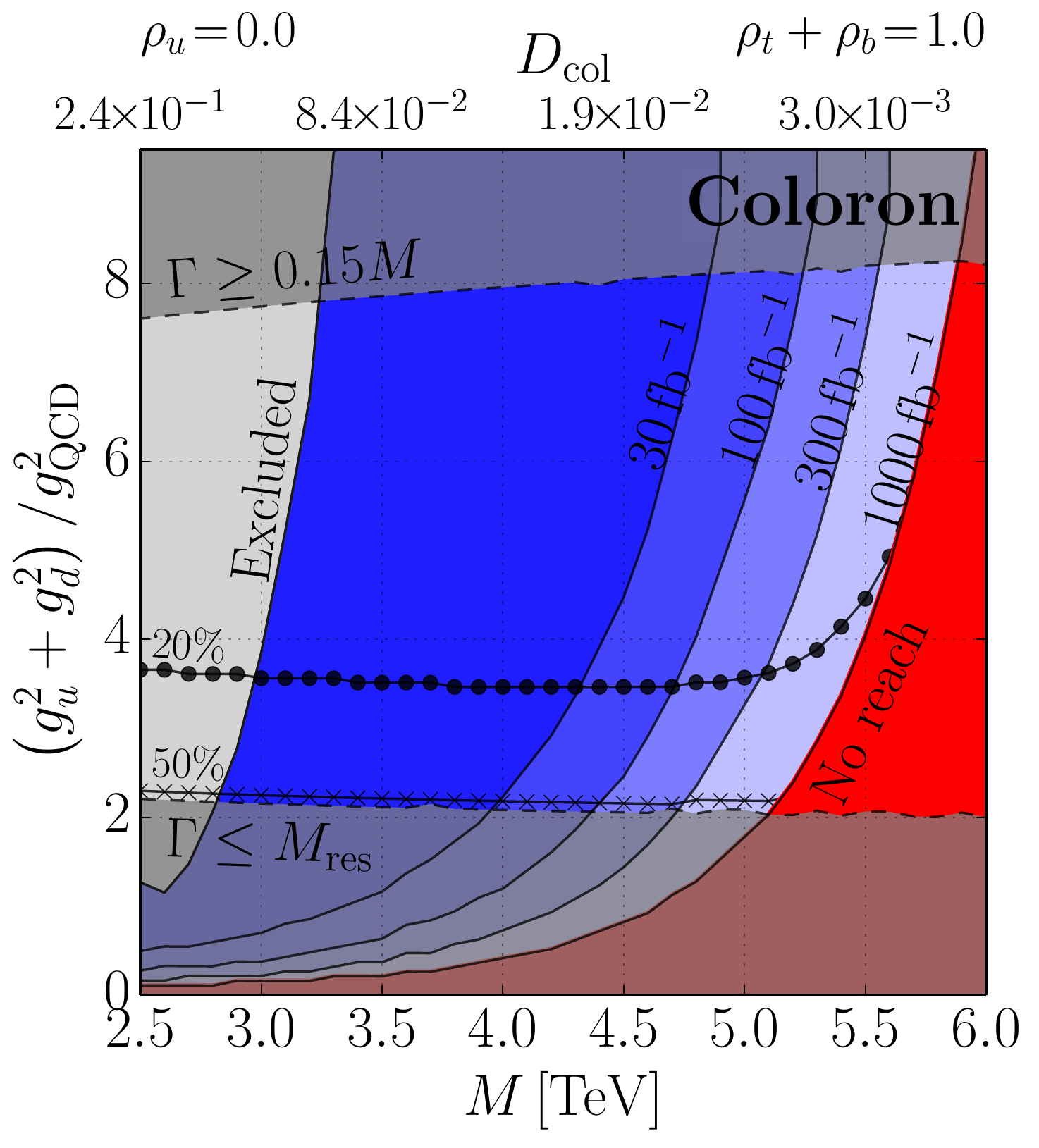}
\includegraphics[width=0.31\textwidth, clip=true]{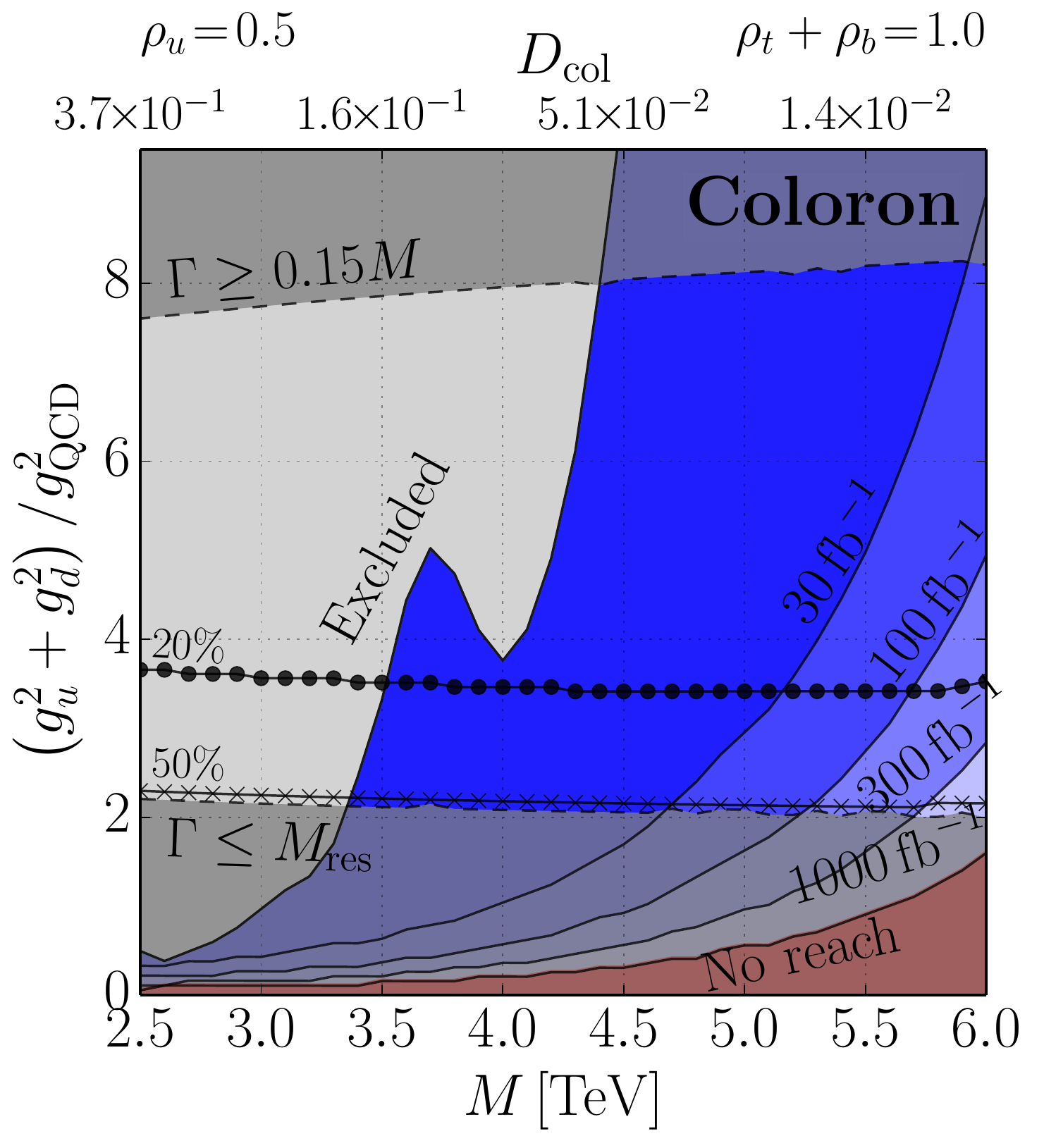}
\includegraphics[width=0.31\textwidth, clip=true]{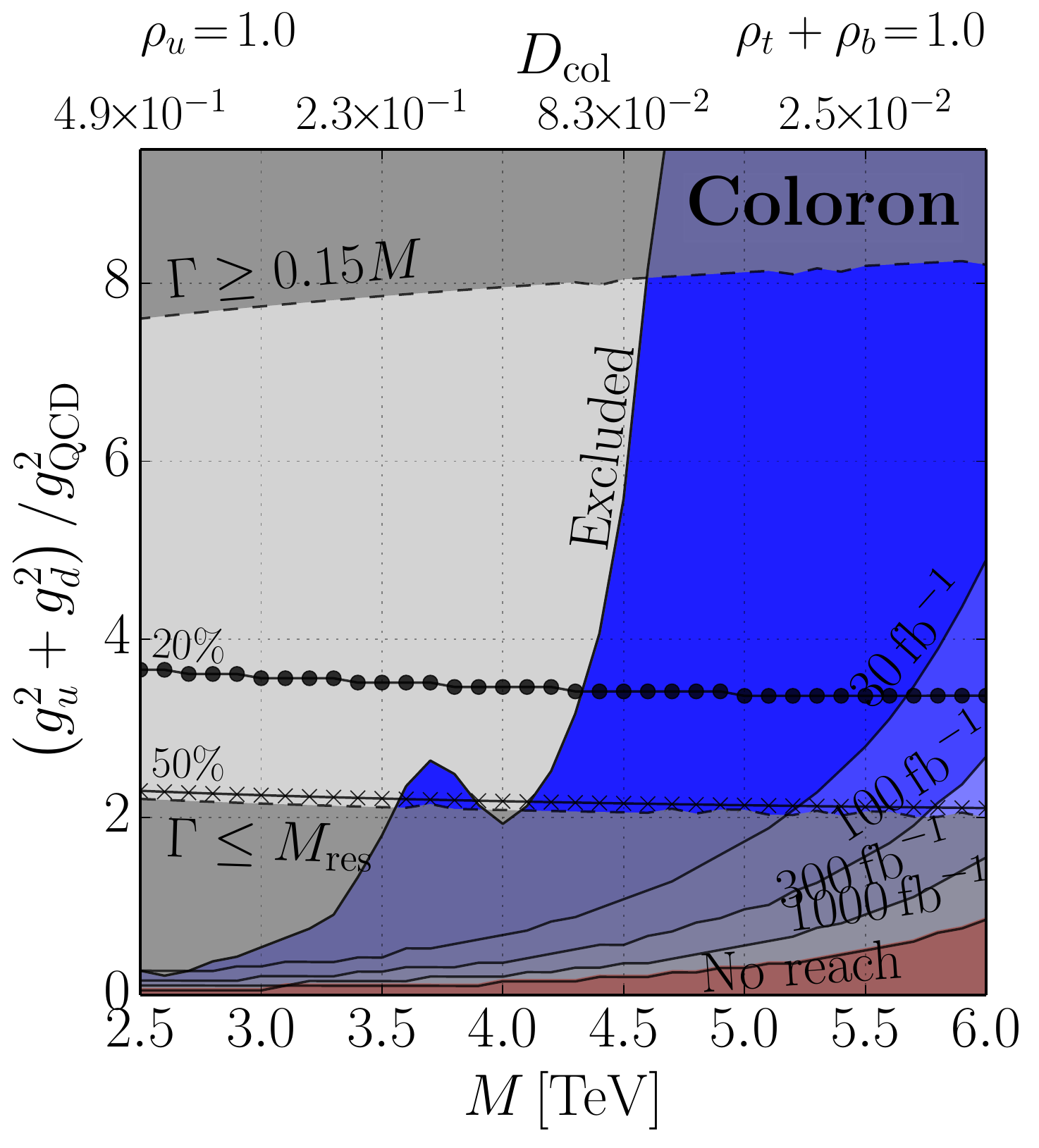}
\\
\includegraphics[width=0.31\textwidth, clip=true]{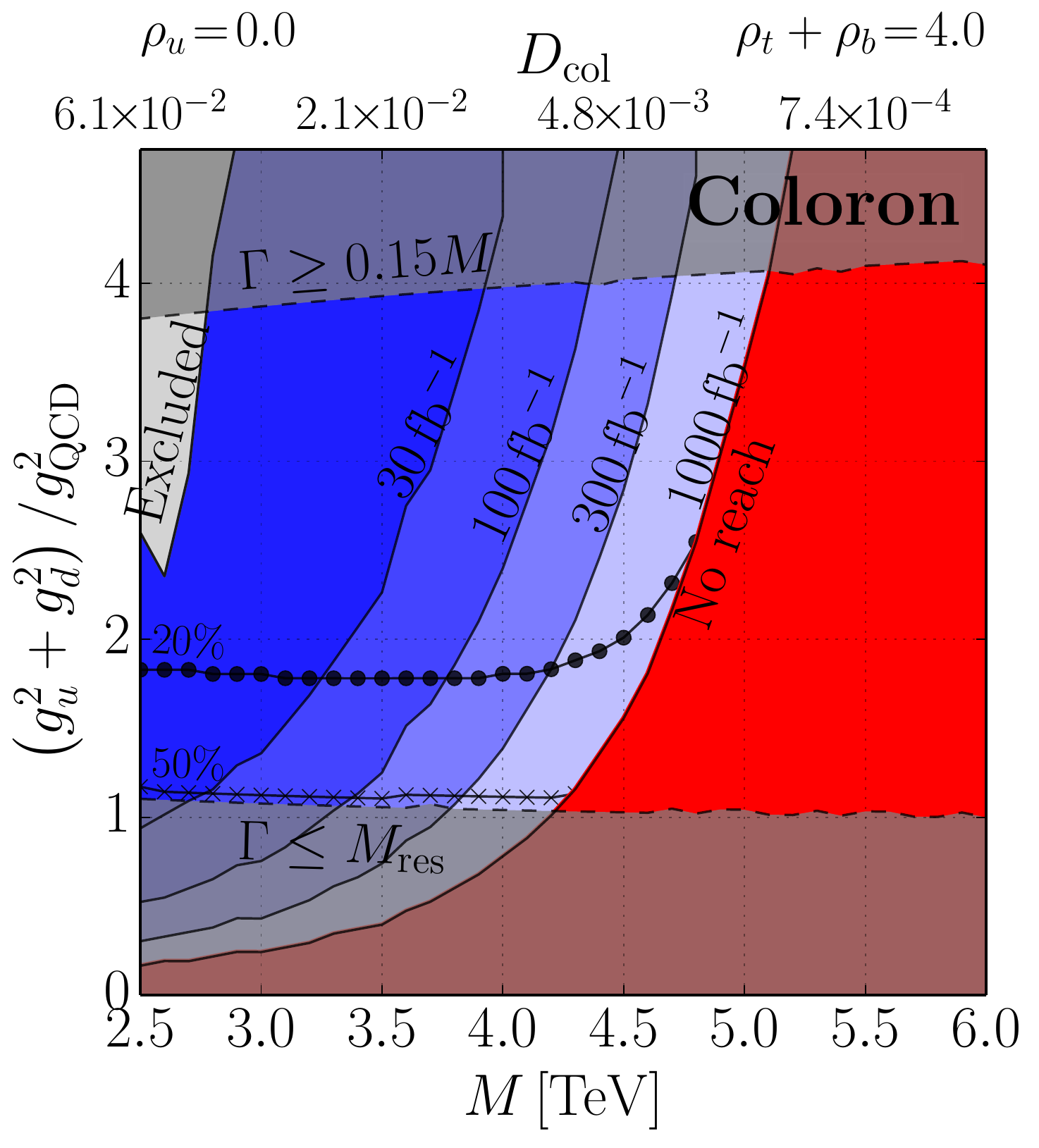}
\includegraphics[width=0.31\textwidth, clip=true]{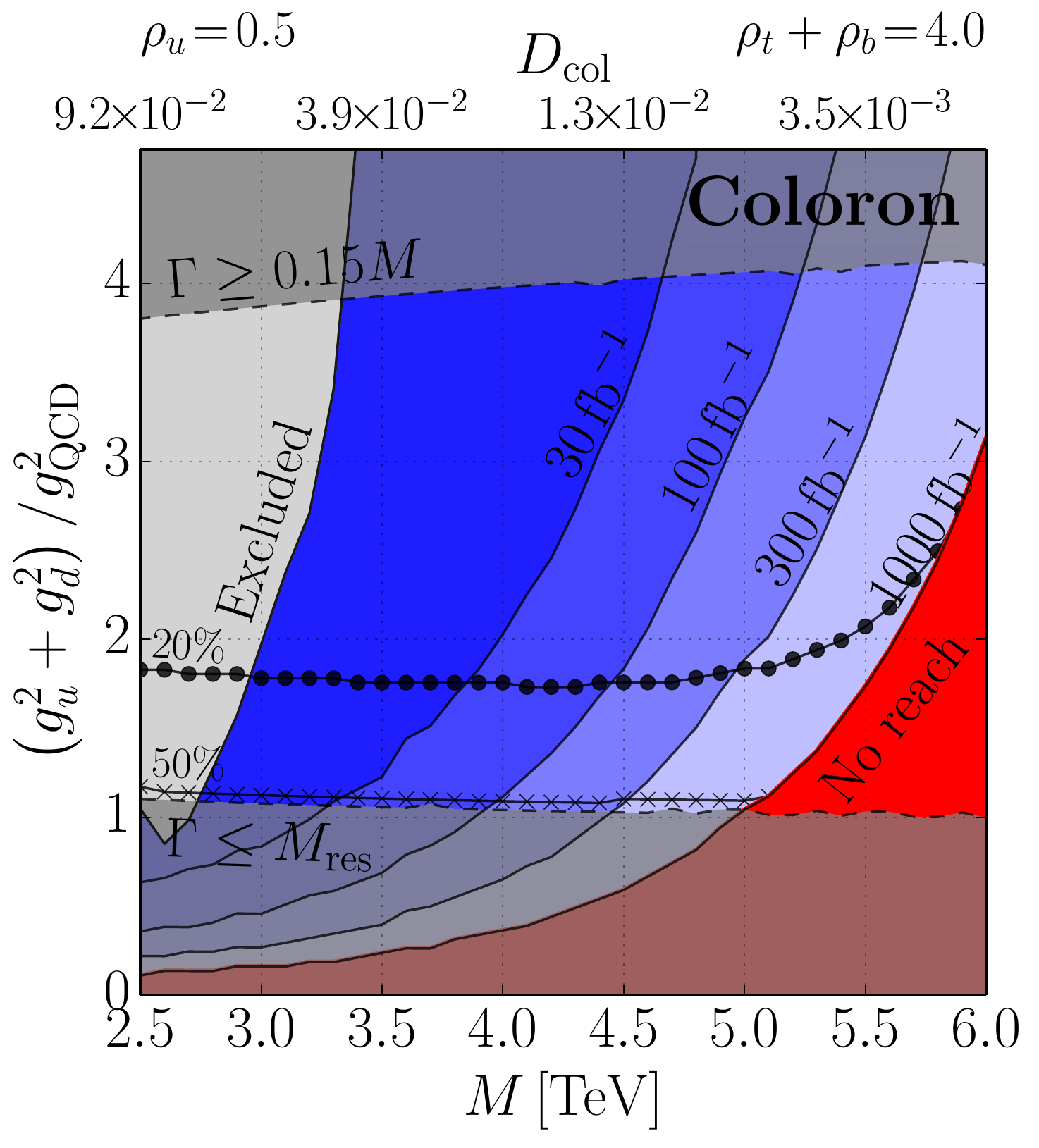}
\includegraphics[width=0.31\textwidth, clip=true]{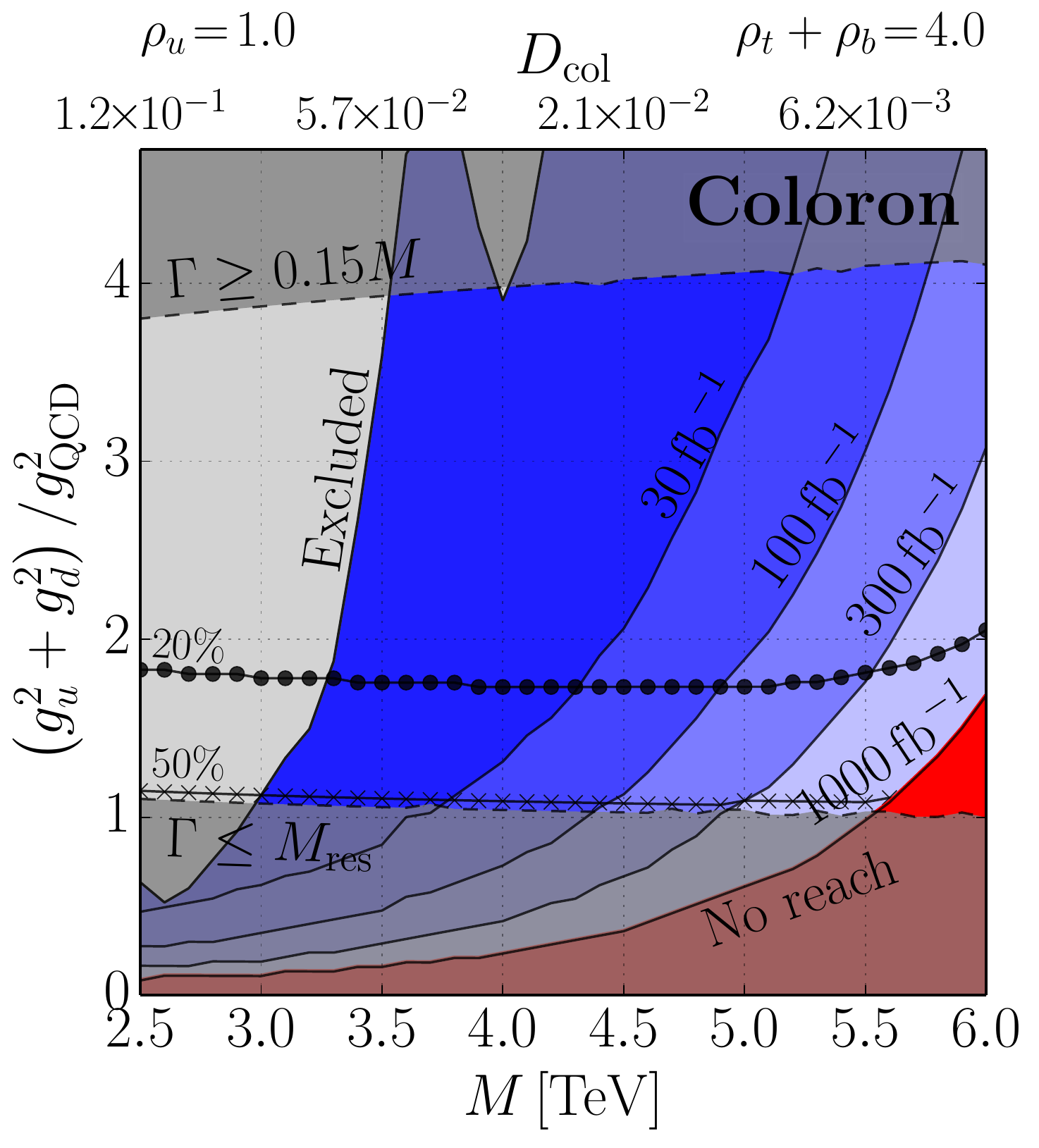}
}
\caption{
Region of parameter space 
where the color discriminant variable analysis applies at the LHC with $\sqrt{s} = 14\,\tev$, for different sets of coupling ratios, $\rho_q \equiv \frac{g_q^2}{g_u^2+g_d^2}$.
The  5$\sigma$ discovery reach, with statistical and systematic uncertainties included, is shown in varying shades of blue for luminosities ranging from 30 $\ifb$ to 1000 $\ifb$. The red area marked ``no reach'' lies beyond the discovery reach at 1000 $\ifb$ The gray area on the left of each plot marked ``excluded'' has been excluded by the 8 $\tev$ LHC~\cite{CMS:kxa}.  In the region above the dashed line marked $\Gamma \ge 0.15M$, the narrow-width approximation used in dijet resonance searches is not valid~\cite{Bai:2011ed, Haisch:2011up, Harris:2011bh}. In the region below the horizontal dashed line marked $\Gamma \le \mres$, the experimental mass resolution is larger than the intrinsic width~\cite{CMS:kxa}, so that one cannot determine $\dcol$. In each figure, for fixed $\rho_q$, $\dcol$ is a function of resonance mass only, with values shown along the upper horizontal axis. The contours marked $20\%$ and $50\%$ indicate the region above which the uncertainty in measuring $\dcol$, as estimated in Sec.~\ref{sec:sens}, is lower than $20\%$ and $50\%$, respectively. 
}
\label{fig:param_space_col}
\end{figure}

\begin{figure}[t]
{
\includegraphics[width=0.31\textwidth, clip=true]{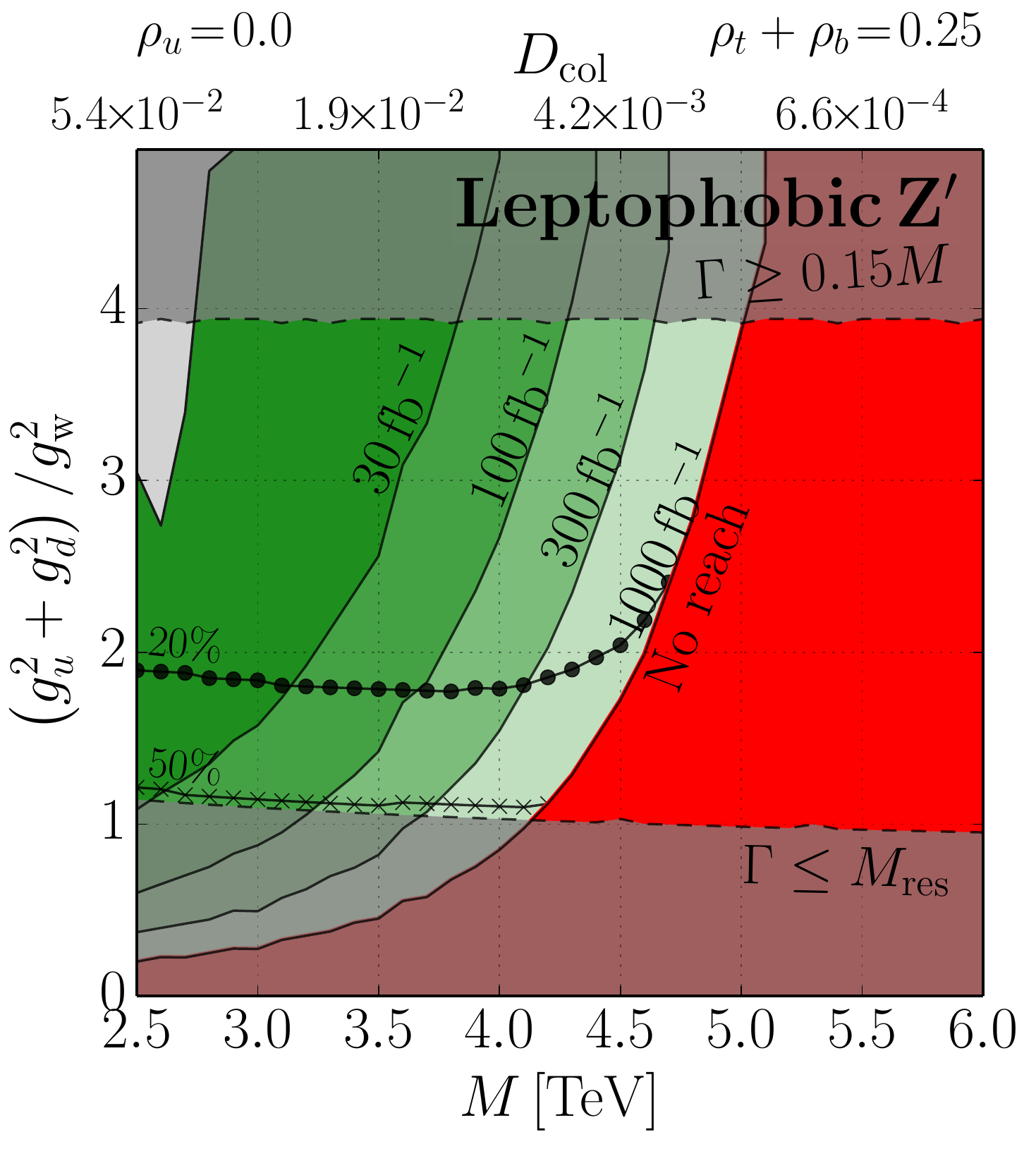}
\includegraphics[width=0.31\textwidth, clip=true]{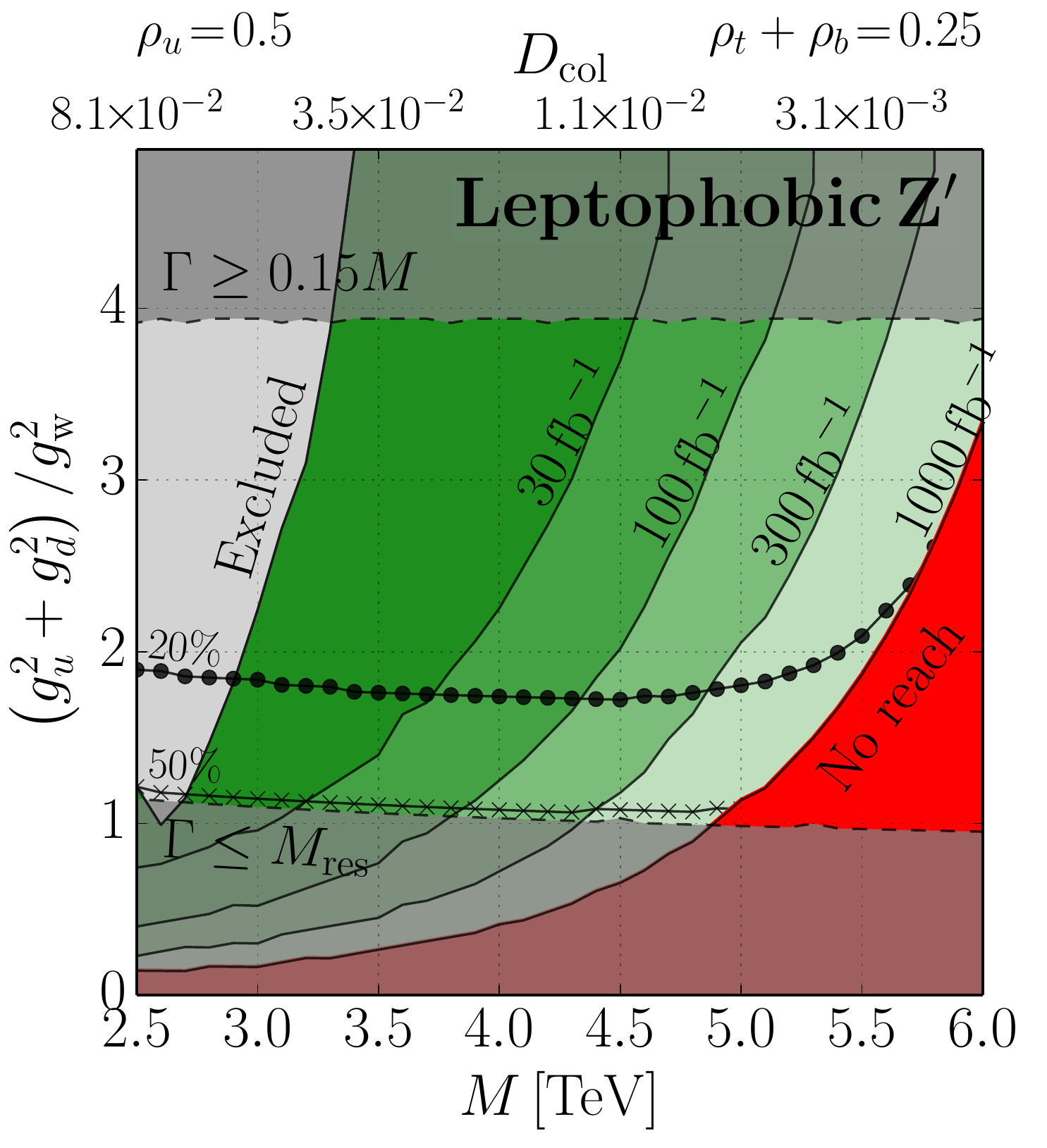}
\includegraphics[width=0.31\textwidth, clip=true]{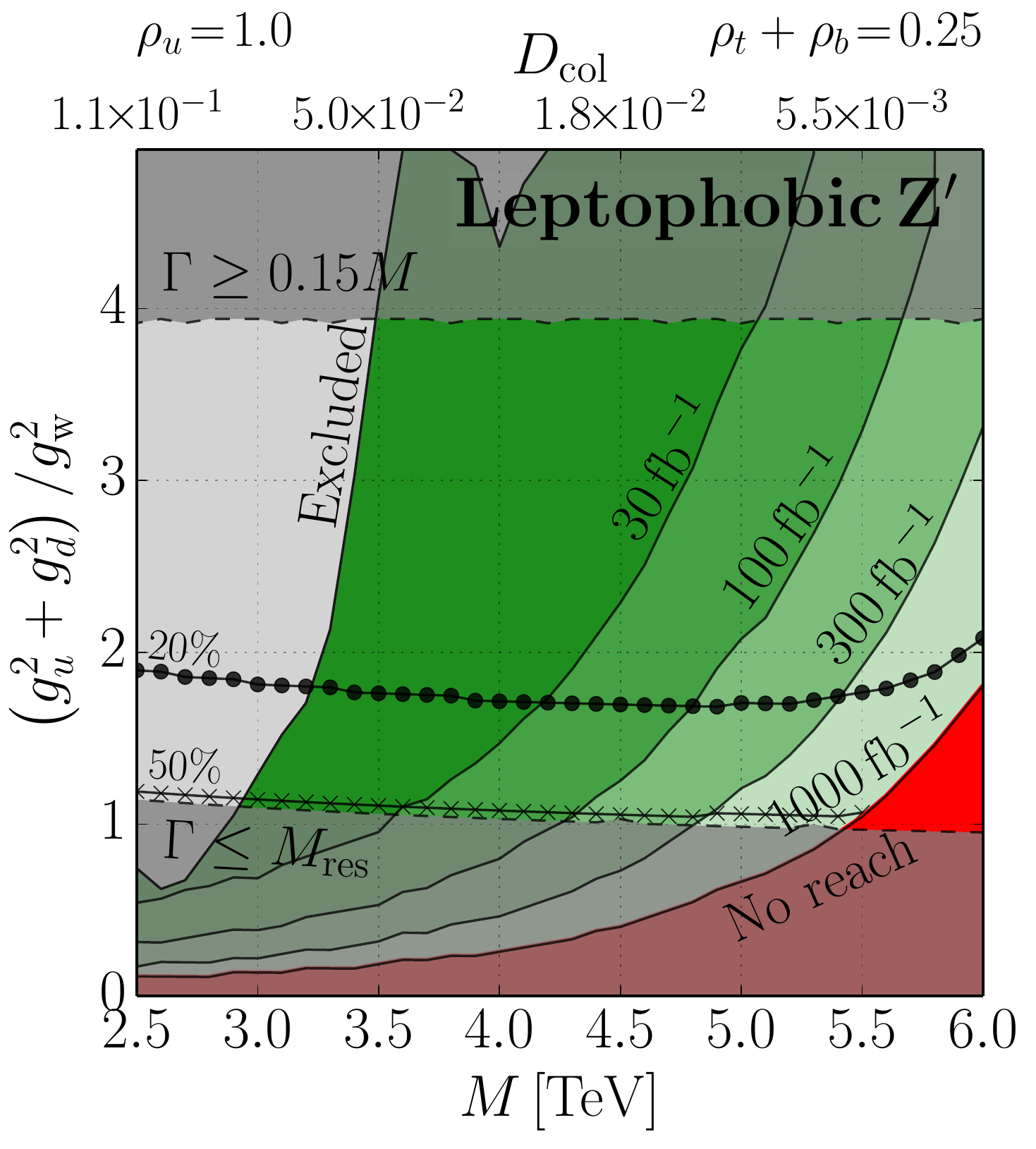}
\\
\includegraphics[width=0.31\textwidth, clip=true]{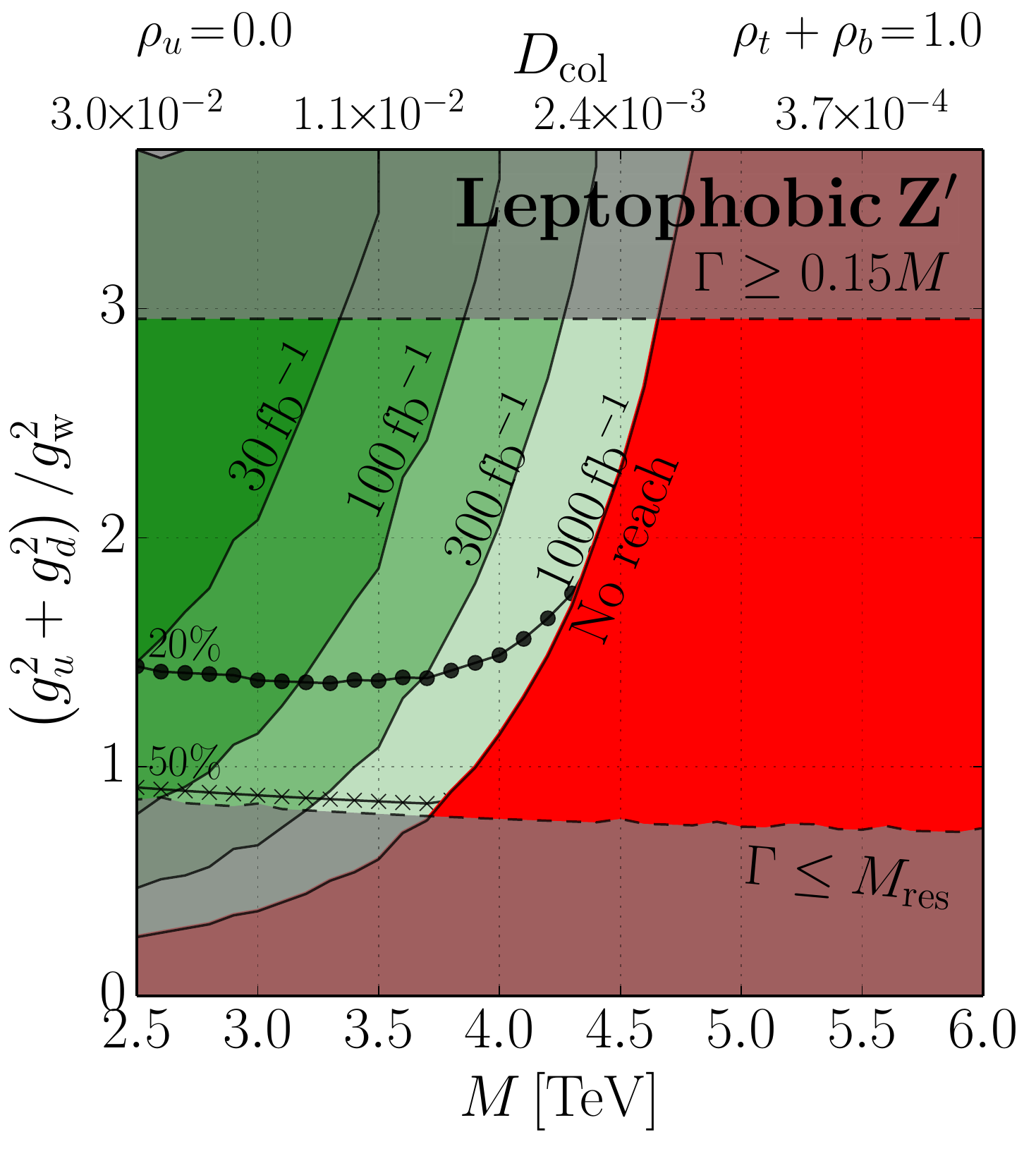}
\includegraphics[width=0.31\textwidth, clip=true]{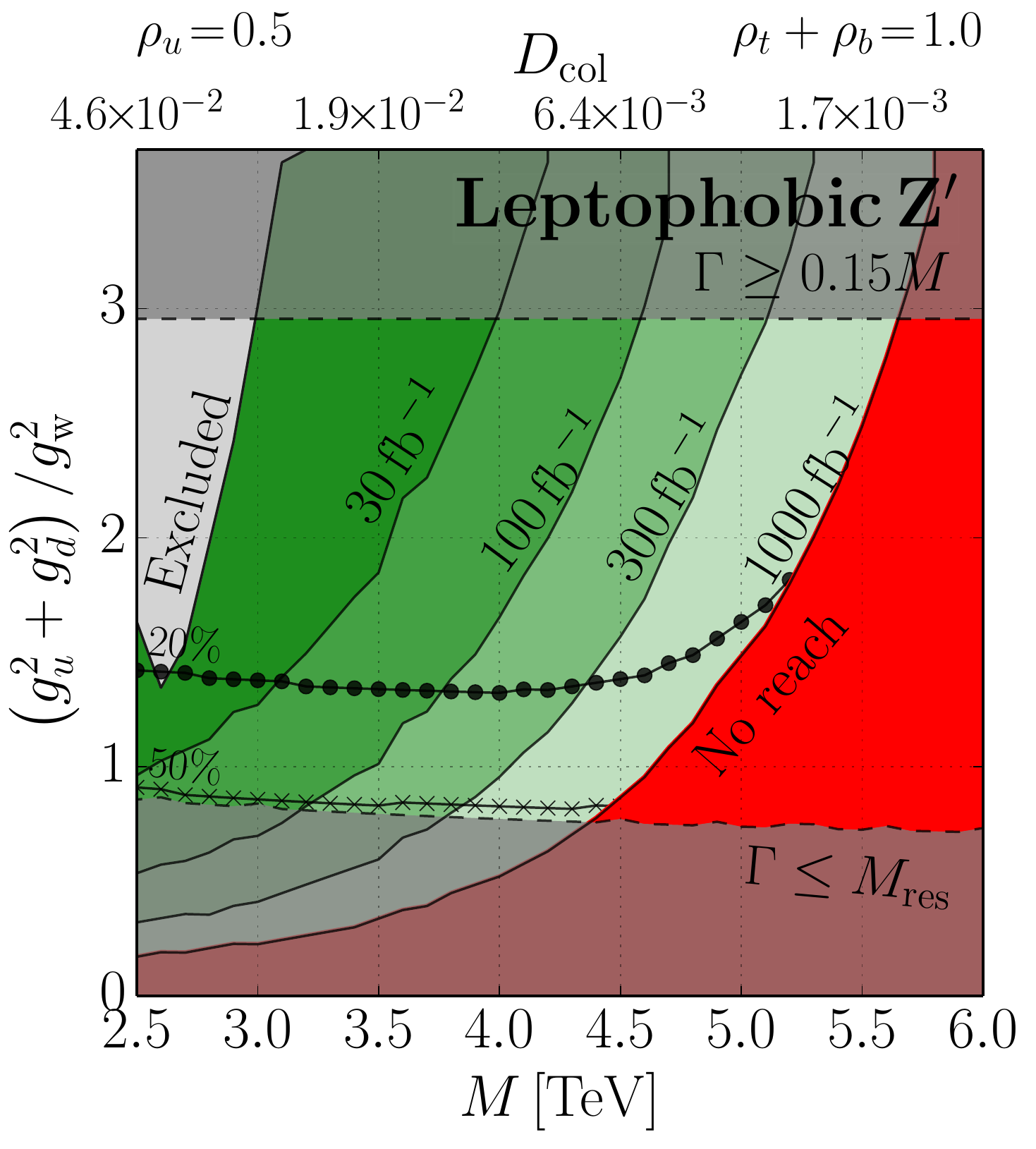}
\includegraphics[width=0.31\textwidth, clip=true]{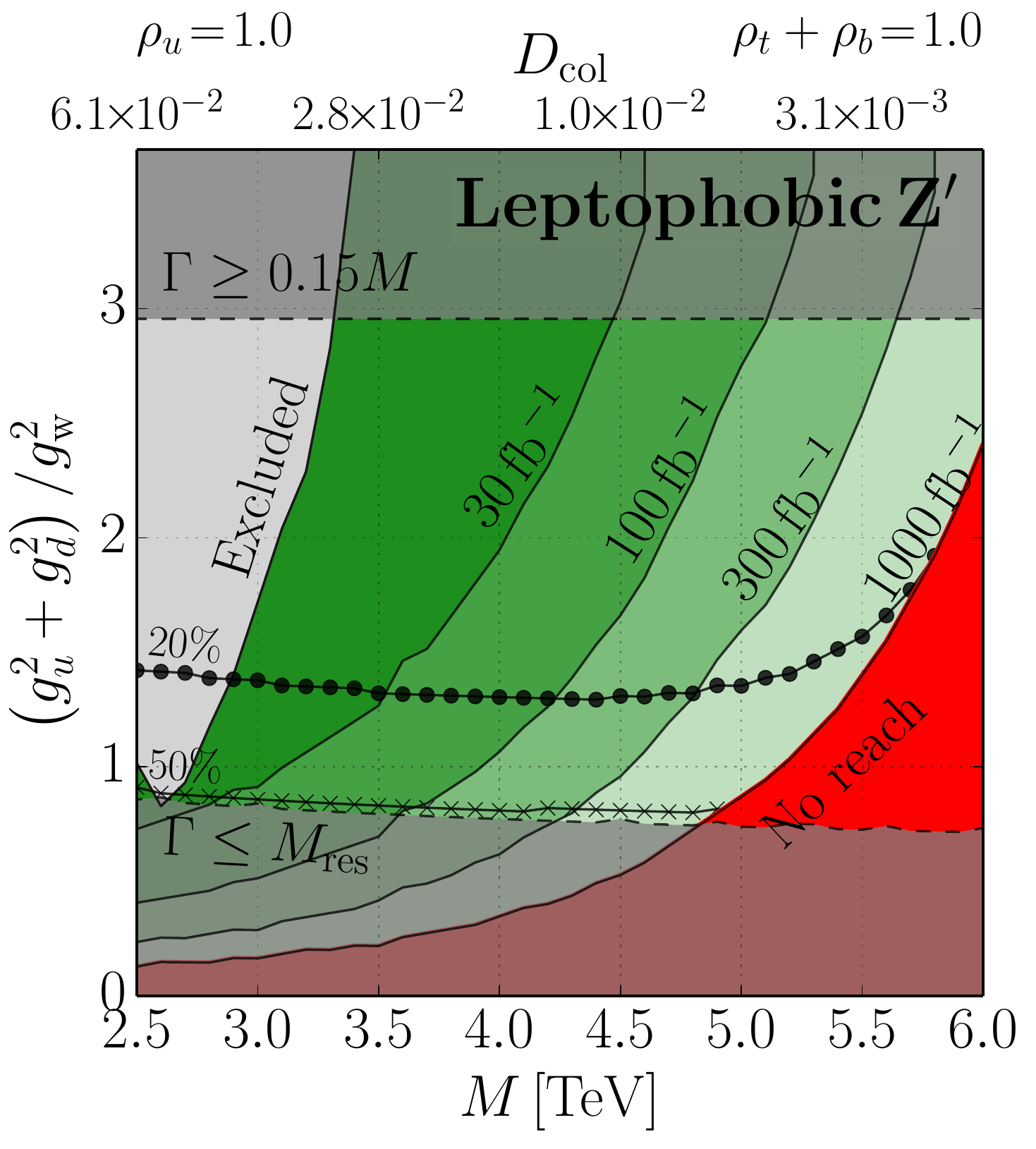}
\\
\includegraphics[width=0.31\textwidth, clip=true]{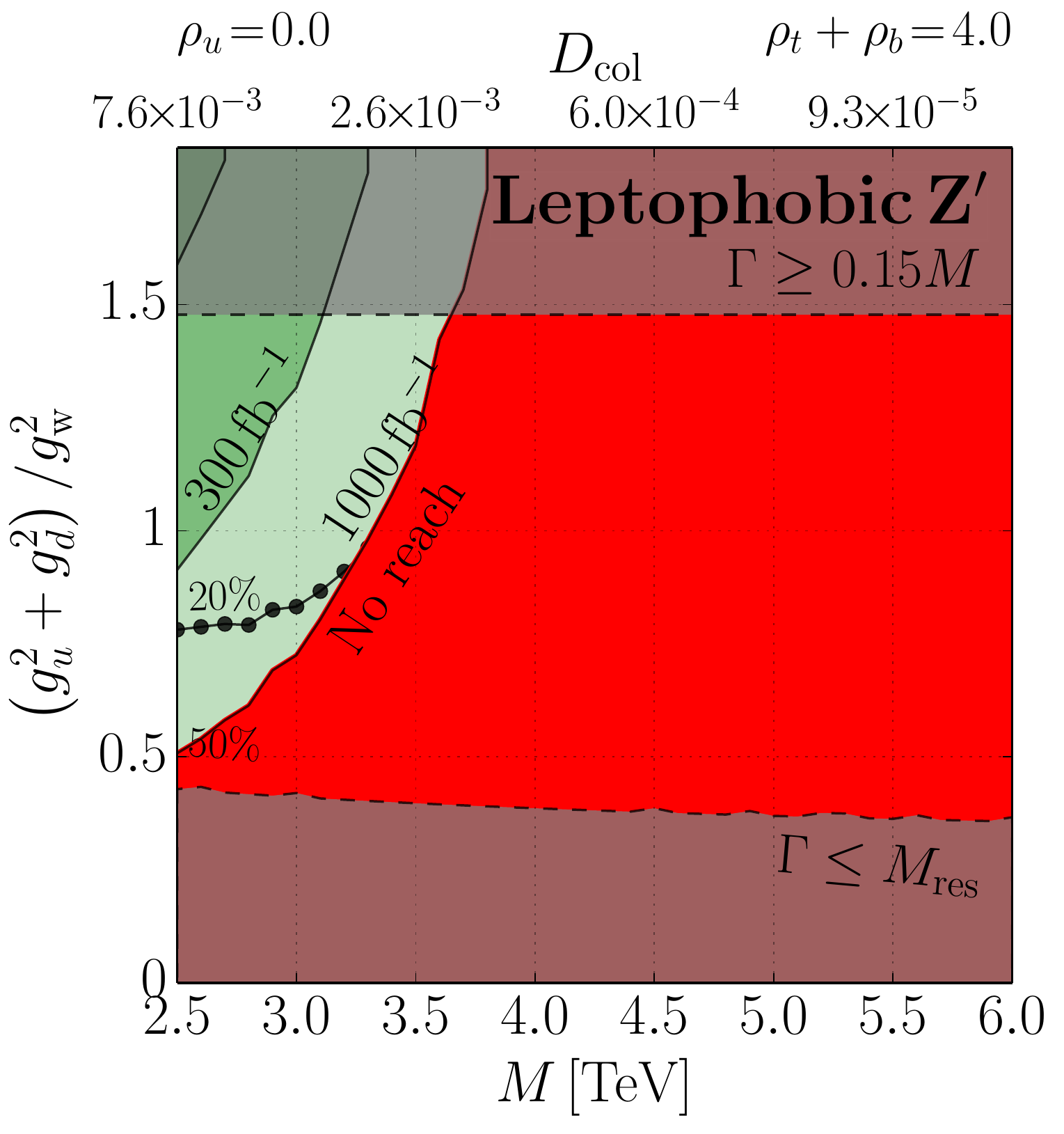}
\includegraphics[width=0.31\textwidth, clip=true]{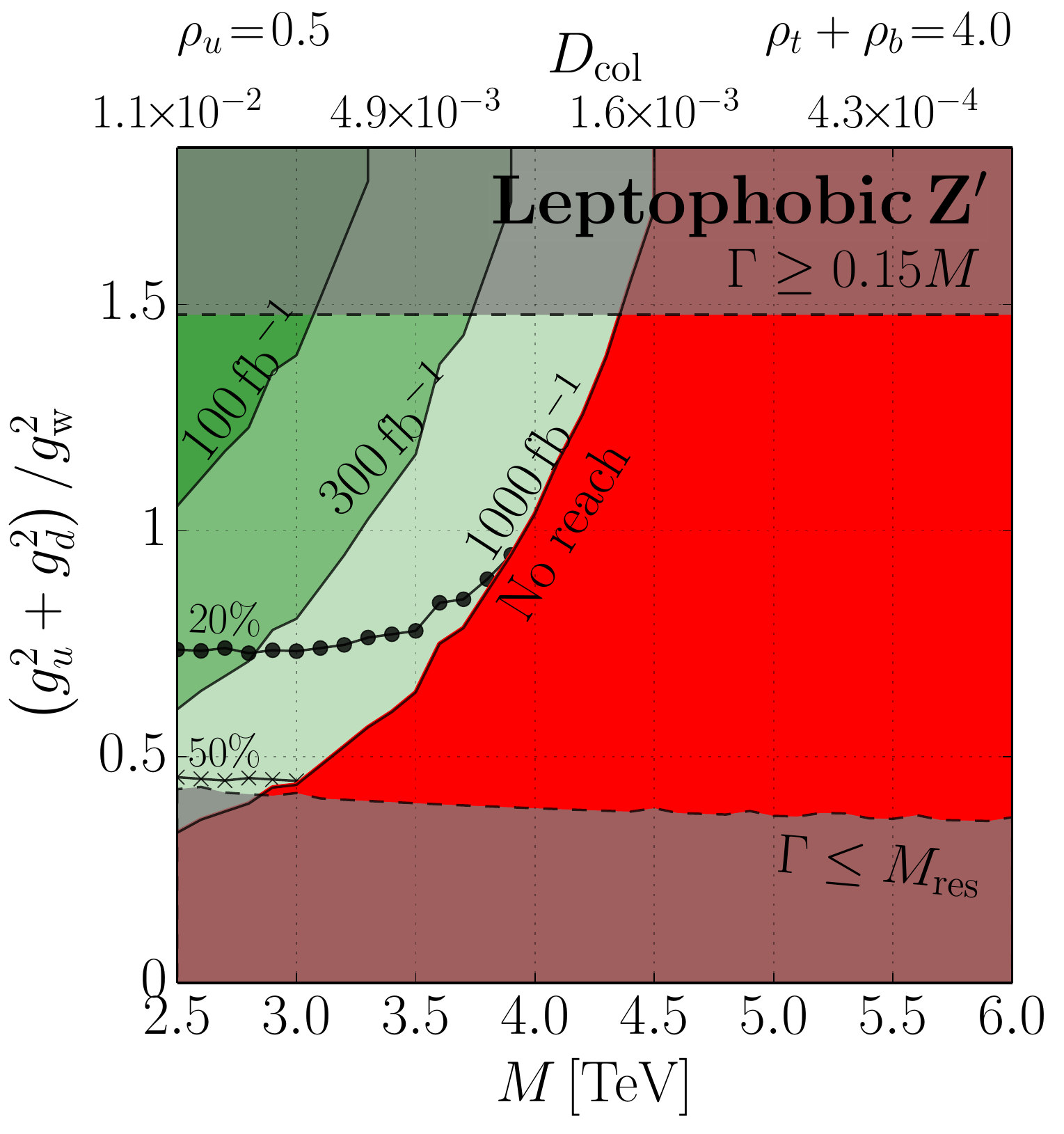}
\includegraphics[width=0.31\textwidth, clip=true]{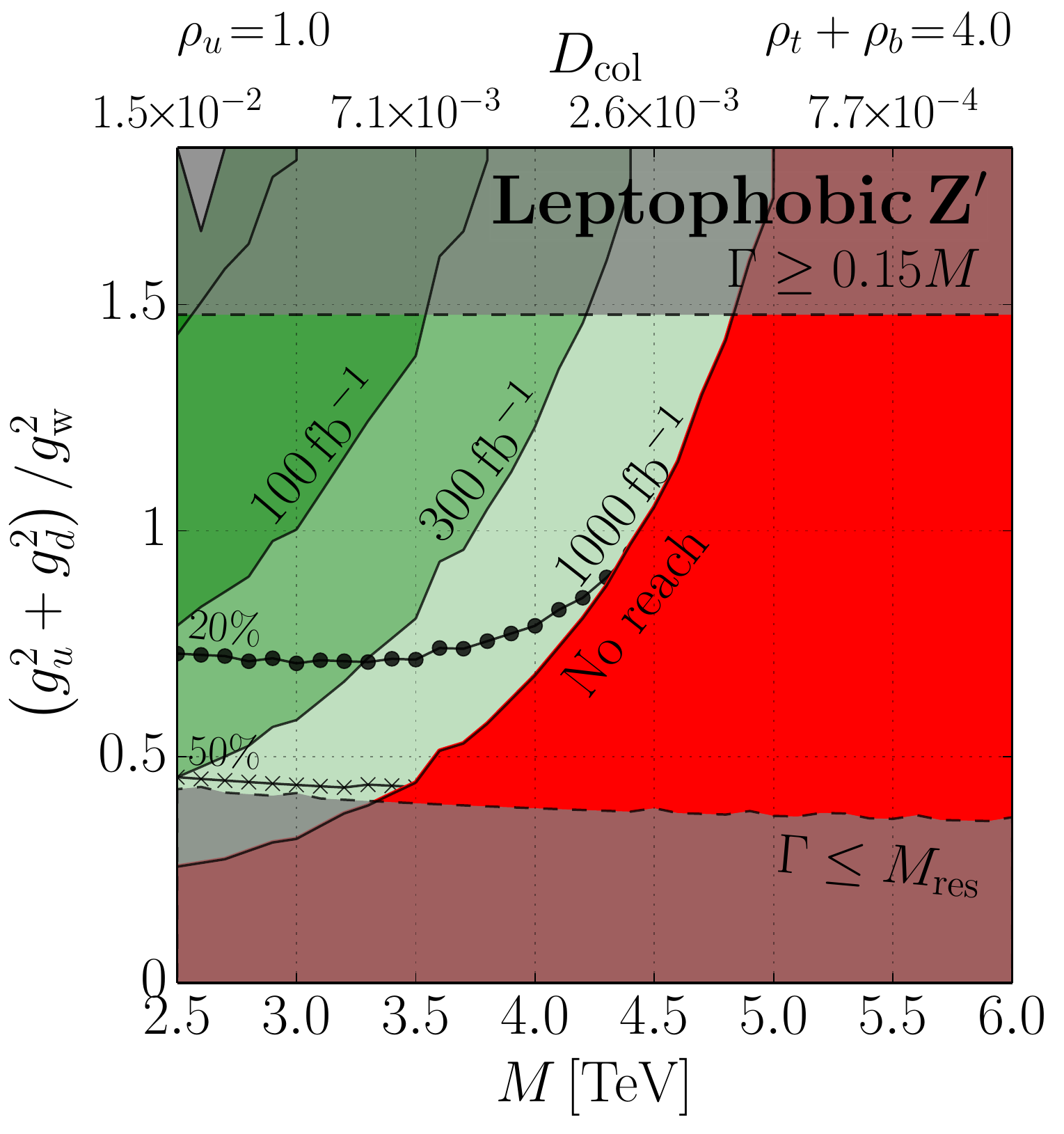}
}
\caption{
Same as Fig.~\ref{fig:param_space_col} but for a flavor-nonuniversal $\zp$. Notice that while a leptophobic $\zp$ having relatively large couplings to quarks in the third generation is still within the reach of the future LHC, its total width would typically be too large to be included in analyses for narrow resonances.
}
\label{fig:param_space_zp}
\end{figure}


\section{Applying the Color Discriminant Variable to Flavor-non-universal Models at the LHC}
\label{sec:result-dcols}
In this section we illustrate how the color discriminant variable $\dcol$ (as described in Section \ref{subsec:flavnon-univ-gen}) may be used to distinguish whether a newly discovered   dijet resonance is a coloron or a leptophobic $\zp$ even if it is flavor non-universal.  As previously mentioned, we will focus on resonances having masses of $2.5-6\,\tev$ at the $\sqrt{s} = 14\,\tev$ LHC with integrated luminosities up to $1000\,\ifb$. The values of $\dcol$ as well as other observables have been evaluated using the uncertainties estimated in Section \ref{sec:sens} and the region of parameter space to which this analysis is applicable was identified in Section \ref{sec:paramspace}.

\subsection{Demonstration that $C$ and $\zp$ lie in different regions of coupling ratio space}

As we have seen, the value of $\dcol$ at a fixed mass and dijet cross section may correspond to a variety of combinations of values of the three ratios of couplings, the up ratio ($\frac{g_u^2}{g_u^2+g_d^2}$), the top ratio ($\frac{g_t^2}{g_u^2+g_d^2}$), and the bottom ratio ($\frac{g_b^2}{g_u^2+g_d^2}$). The last two are directly determined from the measurements of $\sigma_{t\bar{t}}$ and $\sigma_{b\bar{b}}$ while $\dcol$ is relatively insensitive to the first ratio, as mentioned in Section \ref{subsec:flavnon-univ-gen}. The question is, therefore, whether measuring the mass, width, dijet cross-section, $\sigma_{t\bar{t}}$ and $\sigma_{b\bar{b}}$ can definitively identify the color charge of a newly discovered resonance.  We find that it can.  We will illustrate this finding for resonances of mass $3\,\tev$ and $4\,\tev$, as we have seen in Fig.~\ref{fig:param_space_col} and \ref{fig:param_space_zp}  that most colorons with lower masses are excluded by the current experiments and most $\zp$ bosons with higher masses are not within reach of the future LHC run at $1000\,\ifb$.

In Fig.~\ref{fig:3d_dcol_2_5}, we show the region of parameter space of the three coupling ratios (using the observables $\frac{\sigma_{t\bar{t}}}{\sigma_{jj}}$ and $\frac{\sigma_{b\bar{b}}}{\sigma_{jj}}$ in place of the top and bottom ratios, respectively) in which coloron or $\zp$ models (each displayed as a point) with the same mass lead to a certain range of dijet cross-section and $\dcol$. We choose the range for the dijet cross-section to be within 1 standard deviation of the value that allows a $5\sigma$ discovery at luminosity $1000\,\ifb$. We selected  $\dcol$ to be within $50\%$ of the value $3\times 10^{-3}$ for this illustration as it permits the required measurements to be made for either a coloron or $\zp$ as discussed in Section \ref{sec:paramspace} and Figs~\ref{fig:param_space_col} and \ref{fig:param_space_zp}. Points in the accessible area of parameter space are highlighted in blue if the discoverable resonance is a $C$ and in green if it is a $\zp$. These points lie in the blue regions of Fig. \ref{fig:param_space_col} for colorons or the green regions of Fig. \ref{fig:param_space_zp} for $\zp$.

\begin{figure}[t]
{
\includegraphics[width=0.46\textwidth, clip=true]{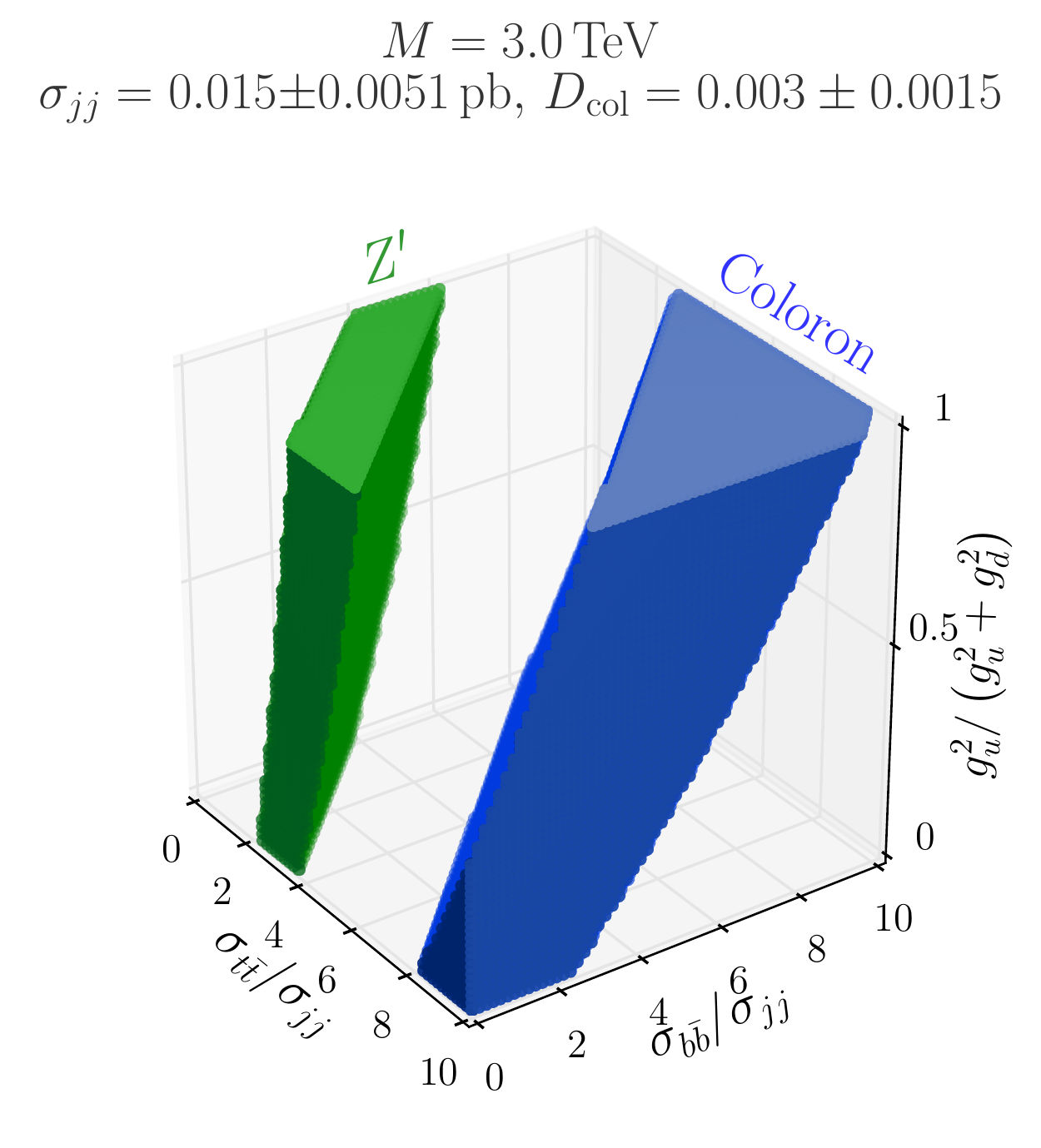}
\includegraphics[width=0.46\textwidth, clip=true]{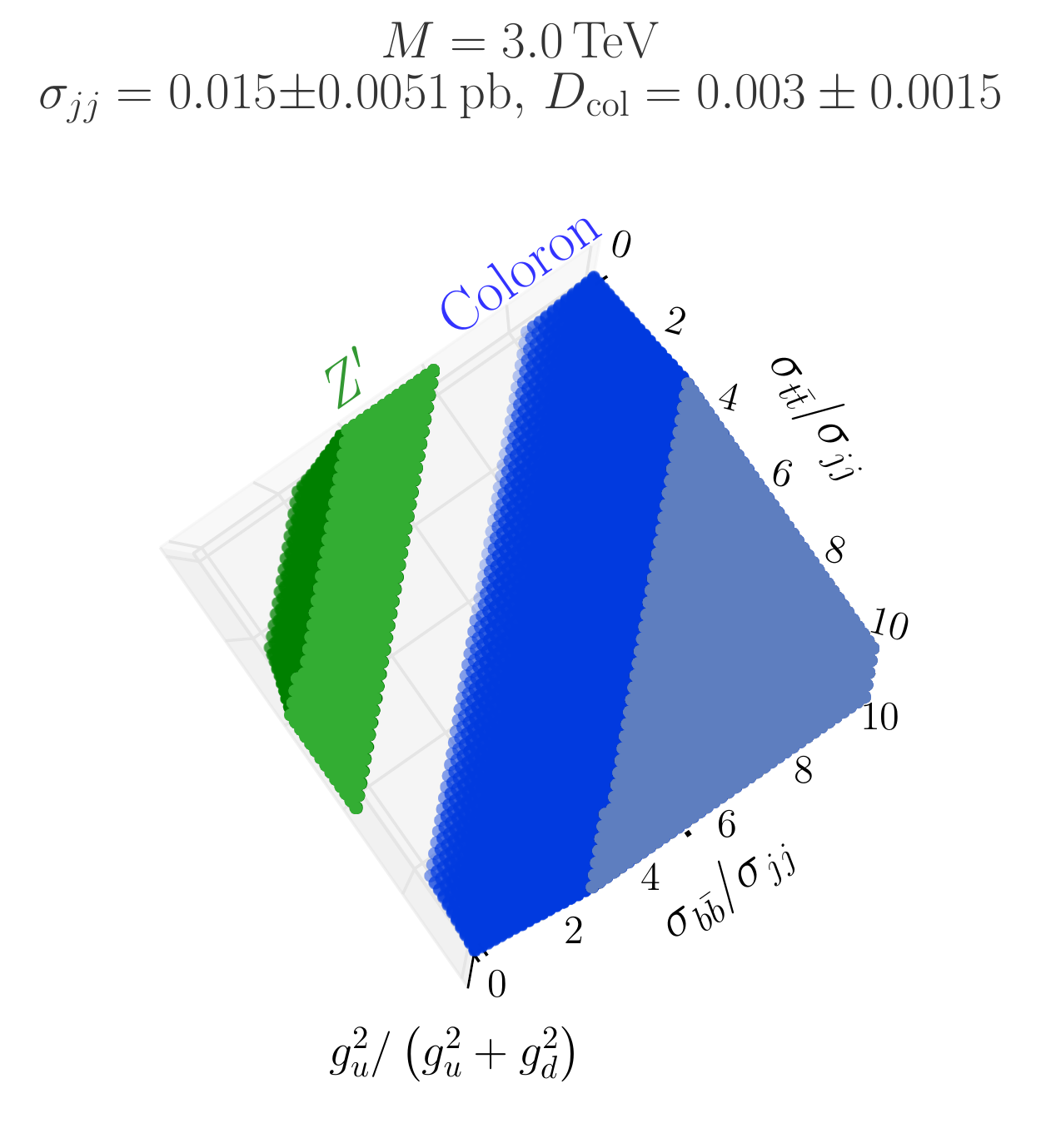}
\\
\includegraphics[width=0.46\textwidth, clip=true]{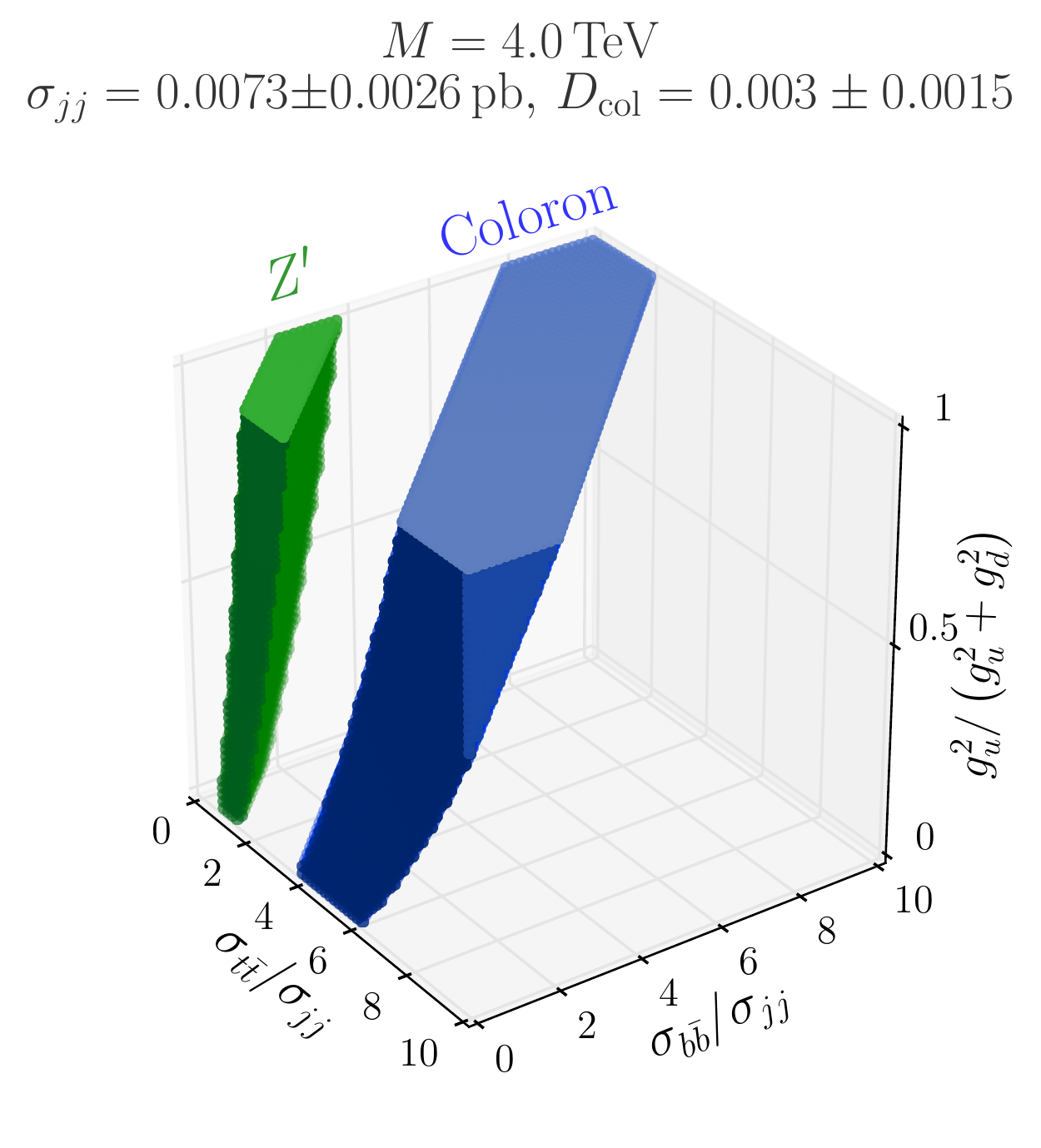}
\includegraphics[width=0.46\textwidth, clip=true]{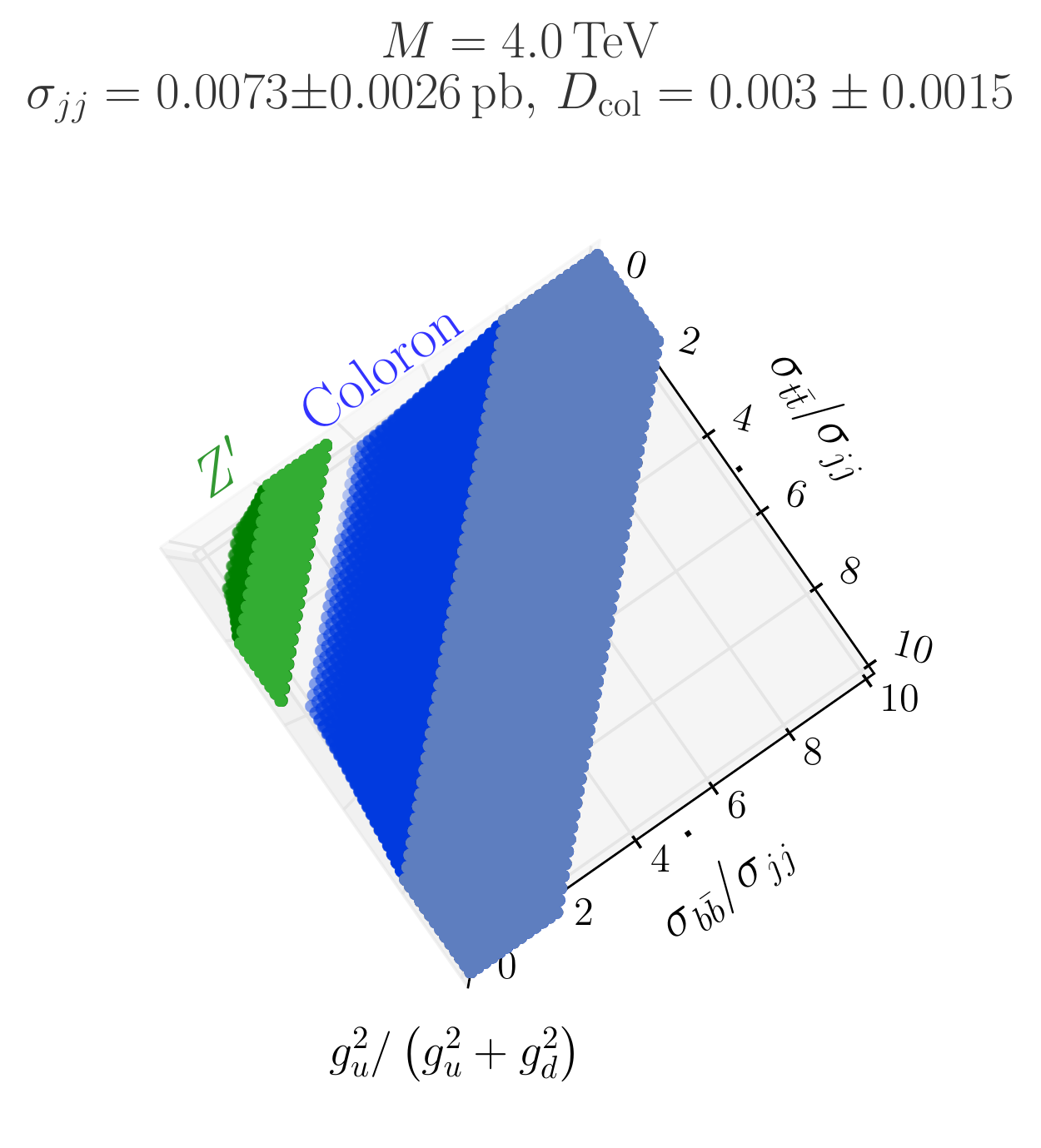}
}
\caption{
Illustration that the proposed measurements suffice to identify the color structure of a new dijet resonance. These plots show regions of the 3-d parameter space $\frac{g_u^2}{g_u^2+g_d^2}$ vs. $\frac{\sigma_{t\bar{t}}}{\sigma_{jj}}$ vs. $\frac{\sigma_{b\bar{b}}}{\sigma_{jj}}$, at fixed values of mass ($3\,\tev$ on the top panels and $4\,\tev$ on the bottom panels) where dijet cross section and $\dcol$ fall within a certain range. The cross section lies within about $35\%$ of the value required for a $5\sigma$ discovery at the LHC with $\sqrt{s} = 14\,\tev$ at $\mathcal{L} = 1000\,\ifb$. Similarly, $D_{col}$ is chosen to lie within $\pm 50\%$ of $3\times 10^{-3}$ for illustration. Points in parameter space that are accessible to the LHC are circles highlighted in blue for $C$ and triangles highlighted in green for $\zp$. Each plot on the left panel is shown again, as viewed from above, on the right. The two views make clear that colorons and Z' bosons lie in distinguishably separate regions of parameter space; in particular, our inability to measure the ratio $\frac{g_u^2}{g_u^2+g_d^2}$ will not prevent us from determining the color structure of a new vector resonance at the $14\,\tev$ LHC.
}
\label{fig:3d_dcol_2_5}
\end{figure}

We now explore the features that Fig.~\ref{fig:3d_dcol_2_5} exhibits. The 3-dimensional plots in the left panel show that models of $C$ and leptophobic $\zp$ which correspond to measurable observables appear in different region of $\frac{\sigma_{t\bar{t}}}{\sigma_{jj}} \,\mathrm{vs}\, \frac{\sigma_{b\bar{b}}}{\sigma_{jj}} \,\mathrm{vs}\,  \frac{g_u^2}{g_u^2+g_d^2}$ three-dimensional parameter space. The symmetry between the $\frac{\sigma_{t\bar{t}}}{\sigma_{jj}}$ and $\frac{\sigma_{b\bar{b}}}{\sigma_{jj}}$ axes illustrates the rarity of having a heavy resonance produced via $b\bar{b}$ annihilation - the only process in which $b$ quarks could contribute to $\dcol$ without a corresponding contribution from $t$ quarks. In addition, we see that while the up ratio is experimentally inaccessible, the top view figures displayed in the right panels show that this does not typically lead to confusion between a coloron and a $\zp$, just  as we have argued in Sec. \ref{subsec:flavnon-univ-gen}. That is, even having projected the 3D data for all up ratio values onto the bottom-ratio vs. top-ratio plane, the blue coloron and green $\zp$ points still lie in distinct regions.

\subsection{Using heavy flavor measurements to tell $C$ from $\zp$}

Fig.~\ref{fig:3d_dcol_2_5} not only shows that the coloron and $\zp$ lie in different regions of parameter space, but also implies that measurements of the top and bottom decays  of the new resonance will almost always enable us to determine its color structure.
Let us assume that a new resonance has been found and that its mass, dijet cross-section, and $\dcol$ have been measured. For illustration, take the values of these observables to be those used in Fig.~\ref{fig:3d_dcol_2_5} (for the same LHC energy,  luminosity and estimated uncertainties). Because there is a gap between the $\zp$ and $C$ regions of that figure when the up ratio is $0$, and because the boundaries of the regions are angled rather than vertical, we can see that a $\zp$ with the minimum up ratio value of $0$ would not be mistaken for a coloron.  On the other hand, a $\zp$ with the maximum up ratio (equal to $1$) lies as close as possible to the coloron region of parameter space. So, to see how close the two regions can get, we will want to compare a $\zp$ with an up ratio of $1$ (one that does not couple to $d$ or $s$ quarks) to a coloron with varying values of the up ratio.

This very comparison is presented in Fig.~\ref{fig:2d_dcol}, which is plotted in the top-ratio vs bottom-ratio plane. The up ratio for the $\zp$ is fixed to be $1$; the up ratio for the $C$ is varied from $0$ (left panels) to $1$ (right panels). We see that as the up ratio for the coloron increases from its minimum to maximum values, the blue coloron region of parameter space moves out from the origin, away from the green $\zp$ region of parameter space. Correspondingly, if we were to decrease the up ratio of the $\zp$ boson, the green $\zp$ region would shift closer to the origin, away from the blue coloron region. In general, the coloron and $\zp$ regions do not overlap.

Given the shape and orientation of the regions corresponding to color-singlet and color-octet resonances in the plots, measuring both the top ratio and bottom ratio would clearly allow us to distinguish the new resonance's color structure. Moreover, we see that if either the top ratio or bottom ratio were measured to be sufficiently large, we would know that the resonance must be a coloron (because the $\zp$ region is already at its maximum distance from the origin). For example, a measurement of $\sigma_{t\bar{t}}/\sigma_{jj} \gtrapprox 6$ for a $3\,\tev$ resonance or $\sigma_{t\bar{t}}/\sigma_{jj} \gtrapprox 3$ for a $4\,\tev$ resonance, for the values of $\sigma_{jj}$ used in Fig.~\ref{fig:2d_dcol}, would identify it as a color-octet.

We note that there could still be a rare situation where our inability to measure the up ratio would prevent us from determining the color structure of a new resonance. The regions of parameter space corresponding to the extreme cases of a coloron with only down-type light quark couplings and a $\zp$ with only up-type light quark couplings could potentially overlap. As mentioned in Sec.~\ref{subsec:flavnon-univ-gen}, this is more likely to happen for heavier resonances due to decreasing values of parton distribution functions for down-type light quarks at higher resonance masses. For example, given our estimates of uncertainties, such an overlap could potentially occur for a $4\,\tev$ resonance as illustrated by the close approach of the $\zp$ and coloron bands in the lower left panel of Fig.~\ref{fig:2d_dcol}.

The determination of the color structure of a resonance generally requires measurements of both $\sigma_{t\bar{t}}$ and $\sigma_{b\bar{b}}$. As with the dijet cross sections, systematic uncertainties for these measurements will be obtained after the experiment (at $14\,\tev$) has started.  While the estimate of these uncertainties lies beyond the scope of this article, our result illustrate that measuring the $t\bar{t}$ and $b\bar{b}$ cross sections to an uncertainty of $\mathcal{O}(1)$ still provides significant information. For the purpose of comparing models and estimating the required uncertainties, additional plots illustrating models of coloron and leptophobic $\zp$ that lead to the same range of values of $\dcol$ are presented in Appendix \ref{sec:exact_dcol}.

\begin{figure}[t]
{
\includegraphics[width=0.99\textwidth, clip=true]{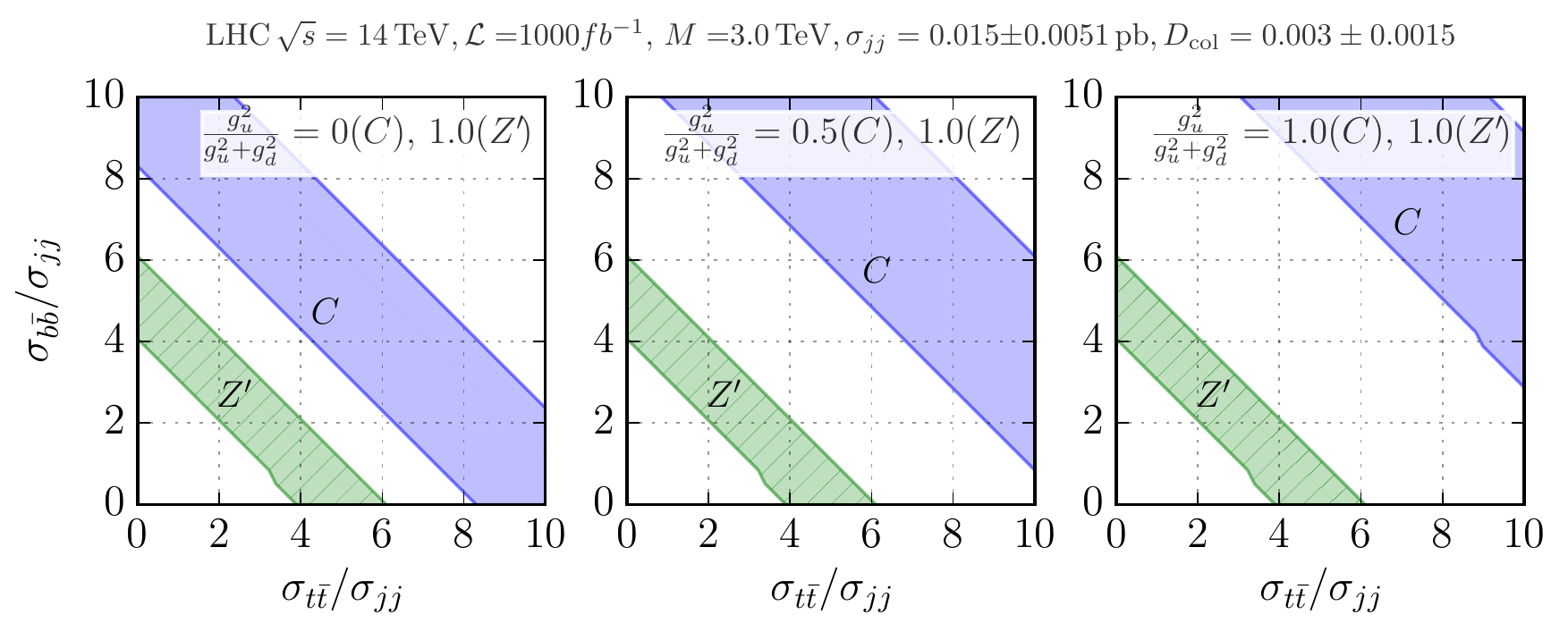}
\\
\includegraphics[width=0.99\textwidth, clip=true]{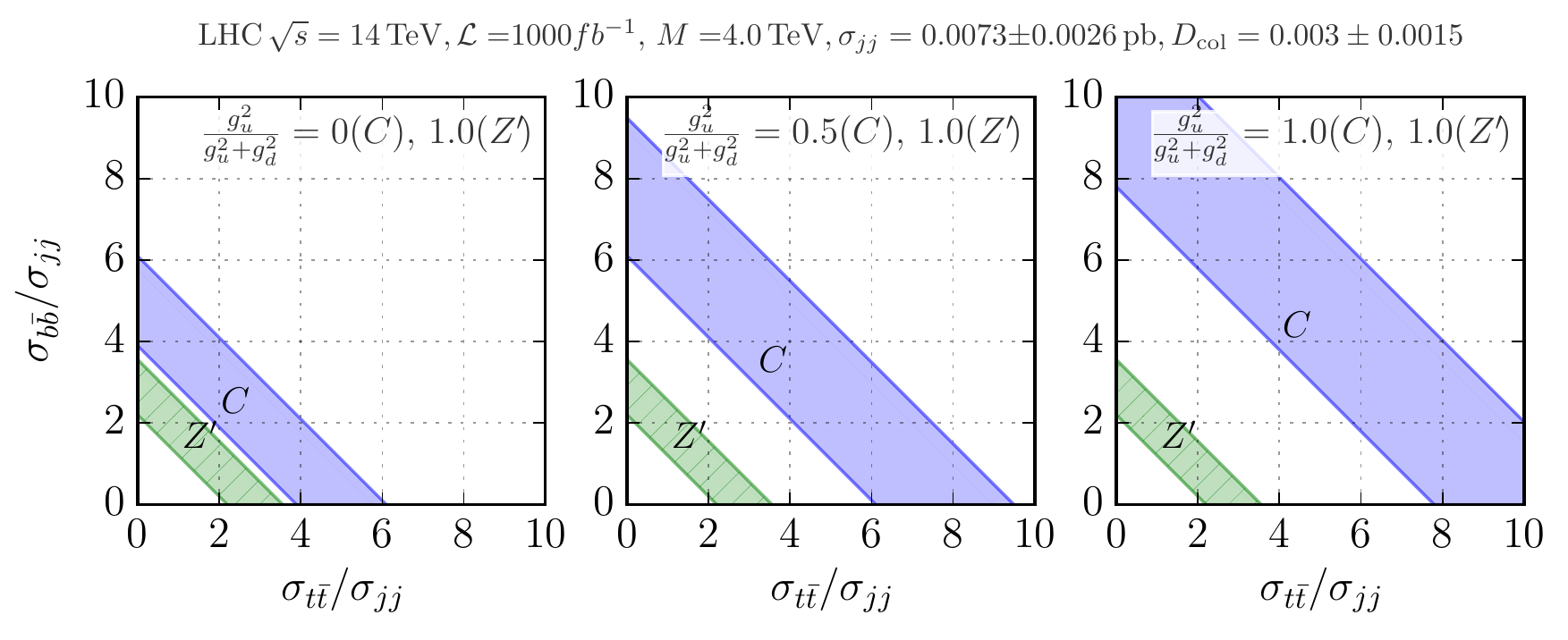}
}
\caption{Further illustration that measuring $\sigma_{t\bar{t}}$ and $\sigma_{b\bar{b}}$ will show whether a resonance of a given mass, dijet cross-section, and $\dcol$ is a coloron or $\zp$. We display regions of parameter space that correspond to a dijet cross section within $1\sigma$ uncertainty of the value required for a discovery at the LHC with $\sqrt{s} = 14\,\tev$ at $\mathcal{L} = 1000\,\ifb$ (the same values used in Fig.\ref{fig:3d_dcol_2_5}), and that also have the illustrative color discriminant variable within a range of $50\%$ around value of $\dcol = 3\times 10^{-3}$, for resonances with masses $3\,\tev$ (top panel) and $4\,\tev$ (bottom panel). Coloron and $\zp$ models that are within reach are displayed in a blue region and a diagonally-hatched green region, respectively.  Note that in nearly all of the displayed areas, the colorons and $Z'$ bosons lie in different regions of parameter space.  However, at the bottom left panel for a 4 TeV resonance, the bands for a $C$ coupling only to down-type light quarks and a $\zp$ coupling only to up-type light quarks approach closely; see also Fig.~\ref{fig:exact_4000_and_4500}.
}
\label{fig:2d_dcol}
\end{figure}

\section{Discussion}
\label{sec:discussions}
The simple topology and large production rate for a dijet final state not only allows a new vector boson to be observed via dijet resonance searches, it also allows the determination of crucial properties of the new resonance. The measurements are particularly useful in distinguishing between a color-octet vector resonance ($C$) and a color-singlet one that couples only to colored particles (leptophobic $\zp$). In this article, we have shown that the method for distinguishing between the two types of resonances in a model-independent manner using the color discriminant variable introduced in Ref.~\cite{Atre:2013mja}, can be extended to more general and realistic scenarios of flavor non-universal couplings in a wide range of models.

The color discriminant variable, $\dcol$, is constructed from measurements available directly after the discovery of the resonance via the dijet channel; namely, its mass, its total decay width, and its dijet cross section. Assuming the new resonance couples identically to quarks of the first two generations, $\dcol$  depends on three model-specific ratios of coupling constants:  the up ratio ($\frac{g_u^2}{g_u^2+g_d^2}$), the top ratio ($\frac{g_t^2}{g_u^2+g_d^2}$), and the bottom ratio ($\frac{g_b^2}{g_u^2+g_d^2}$). We showed that the method is generally not dependent on knowing the up ratio, a quantity which is not presently accessible to experiment. Since $\dcol$ is insensitive to chiral structure, discriminating between color-singlet and color-octet resonances with flavor non-universal couplings requires only measurements of the $t\bar{t}$ and $b\bar{b}$ resonance cross sections. Our results are illustrated in Figs.~\ref{fig:3d_dcol_2_5} and \ref{fig:2d_dcol}, with further scenarios explored in Appendix \ref{sec:exact_dcol}.

Our analysis has assumed that the coloron and $\zp$ have only negligible couplings to any non-Standard Model fermions that may exist. It is straightforward to consider an extension to models where the resonance does couple to new fermions.  While the coloron always has only visible decay channels (as it couples to colored particles), the $\zp$ could have non-negligible decay branching fraction into non-Standard Model invisible particles. In that case, the leptophobic $\zp$ and the coloron would be even easier to distinguish from one another by the color discriminant variable. Simply put, the $\zp$'s invisible decays would increase its total width, which appears in the denominator of the expression for its color discriminant variable%
\footnote{\textit{Cf.} the curly braces of Eq.~(\ref{eq:dcolzp-nonu}).}%
. This means the value of the $\zp$'s color discriminant variable, which is already smaller than that of the coloron by a factor of $8$, would be further reduced due to the appearance of non-negligible invisible decays. Therefore, a leptophobic $\zp$ with invisible decays will correspond to a region in the $\frac{g_u^2}{g_u^2+g_d^2}$ vs. $\frac{\sigma_{t\bar{t}}}{\sigma_{jj}}$ vs. $\frac{\sigma_{b\bar{b}}}{\sigma_{jj}}$ parameter space (such as those presented in Fig.~\ref{fig:3d_dcol_2_5}) that lies even further away from the region occupied by colorons.

To summarize: we have generalized the color discriminant variable for use in determining the color structure of new bosons that may have flavor non-universal couplings to quarks. We focused on resonances having masses $2.5-6.0\,\tev$ for the LHC with center-of-mass energy $\sqrt{s} = 14\,\tev$ and integrated luminosities up to $1000,\,\ifb$. After taking into account the relevant uncertainties and exclusion limits from current experiment and sensitivity for future experiments, we find that the future runs of the LHC can reliably determine the color structure of a resonance decaying to the dijets.

\begin{acknowledgments}
We thank Anupama Atre for her contributions to the previous paper on this topic, upon which the current study is built, and Wade Fisher for discussions on the prospects for measuring $\sigma_{t\bar{t}}$ and $\sigma_{b\bar{b}}$. This material is based upon work supported by the National Science Foundation under Grant No. PHY-0854889. We wish to acknowledge the support of the Michigan State University High Performance Computing Center and the Institute for Cyber Enabled Research. PI is supported by Development and Promotion of Science and Technology Talents Project (DPST), Thailand.
\end{acknowledgments}

\appendix

\section{Uncertainties from PDFs}
\label{sec:uncertainties_due_to_pdf}

We estimate the uncertainty on $\dcol$ due to uncertainties in the parton distribution function, following the authors in \cite{Alekhin:2011sk}, using CT10NLO PDF sets provided by the CTEQ Collaboration \cite{Lai:2010vv}. For an observable $X$,
	\begin{equation}
		\Delta X = \frac{1}{2}\sqrt{\sum_{1}^{N}[X_i^{+} - X_i^{-}]^2}
	\label{eq:delta_x}
	\end{equation}
where $X_i^{+}$ ($X_i^{-}$) is the value calculated using the PDF member corresponding to the ``+'' (``-'') direction of the error. For CT10NLO, $N=26$. The $90\%\,$ C.L. errors from CT10NLO were then rescaled to $68\,\%$ C.L. when presented in our work.

Uncertainties from PDFs contribute to the uncertainty in $\dcol$ via a combination of $W_u + W_c$ and $W_d + W_s$ for the $W_q(M)$ functions defined in Eq.~(\ref{eq:w_function}). In the left panel of Fig.~\ref{fig:wuc_wds_from_ct10nlo}, we plot the central values of $W_u + W_c$ (in red) and $W_d + W_s$ (in blue) with uncertainty bands evaluated as described above. We also provide their fractional errors in the right panel of Fig.~\ref{fig:wuc_wds_from_ct10nlo}.

\begin{figure}[ht]
\includegraphics[width=0.48\textwidth, clip=true]{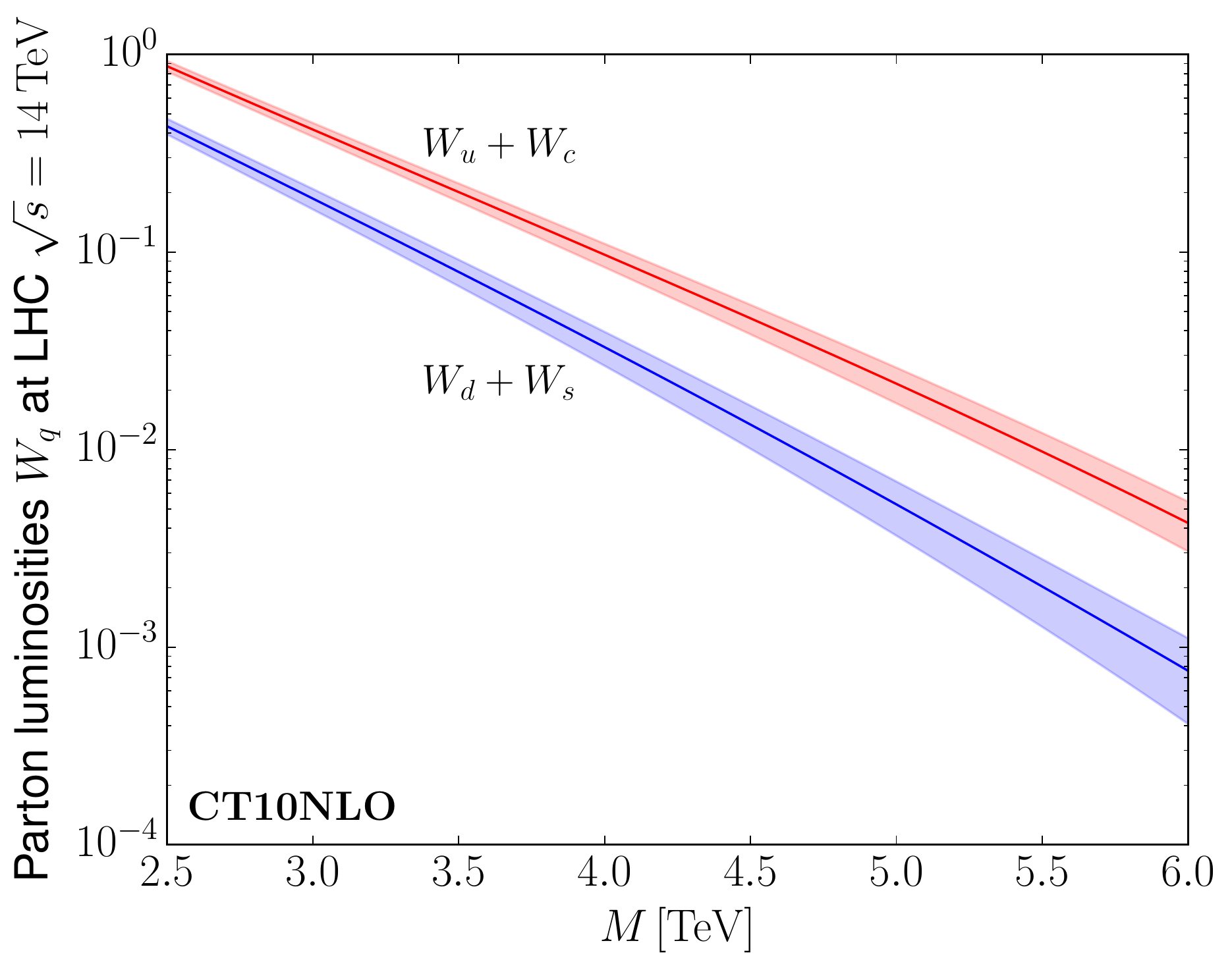}
\includegraphics[width=0.48\textwidth, clip=true]{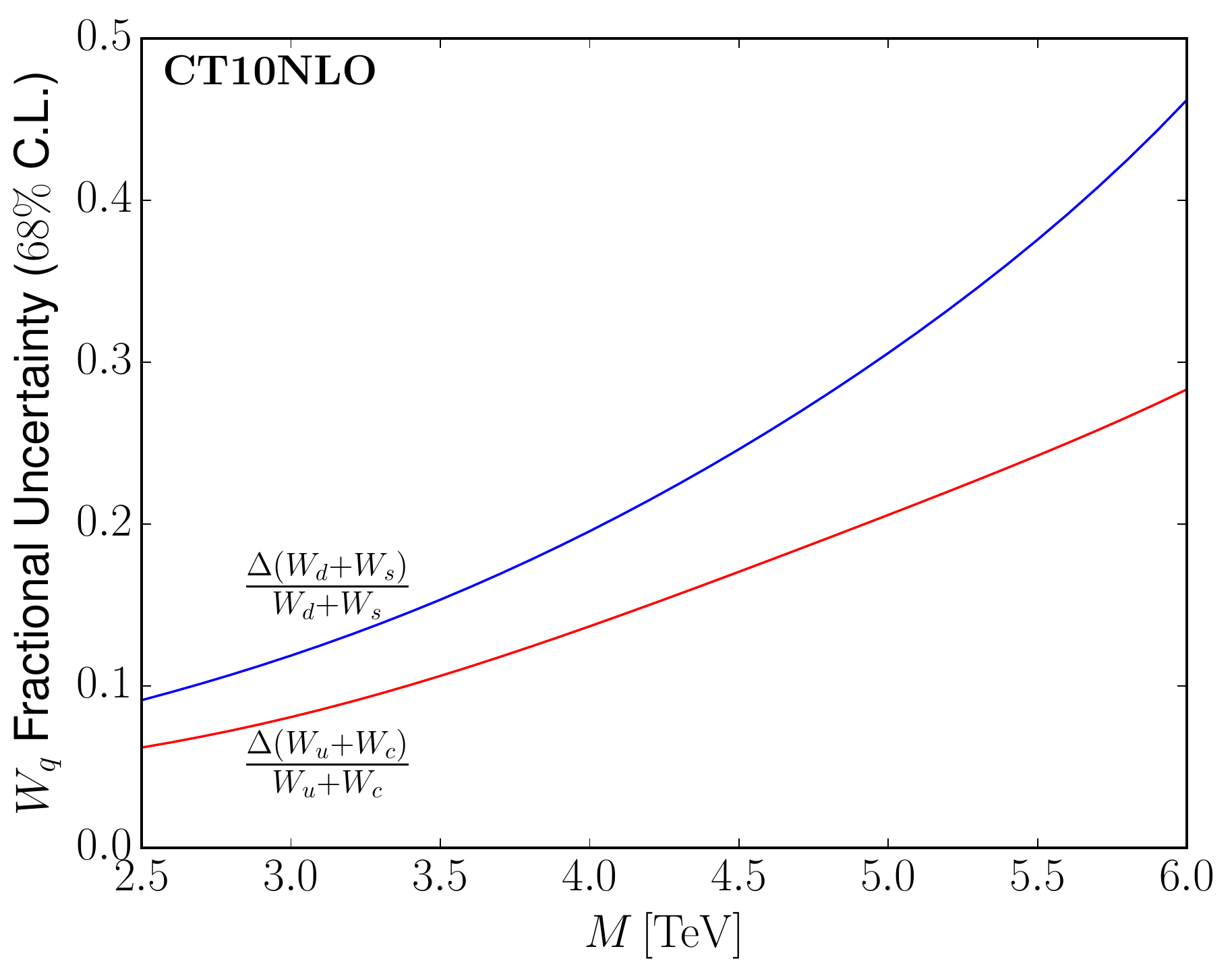}
\caption{(a) Left panel: central values for $W_u + W_c$ (the red band on top) and $W_d + W_s$ (the blue band at the bottom) with uncertainty bands evaluated using CT10NLO PDF set. (b) Right panel: fractional errors for $W_u + W_c$ (red line on top) and $W_d + W_s$ (blue line at the bottom) evaluated using CT10NLO PDF set.
}
\label{fig:wuc_wds_from_ct10nlo}
\end{figure}

\section{$\dcol$ for Viable Models}
\label{sec:exact_dcol}

This appendix illustrates that the proposed measurements of the color discriminant variable ($\dcol$) and resonance cross sections including decays to heavy flavors ($\frac{\sigma_{t\bar{t}}}{\sigma_{jj}}$, and $\frac{\sigma_{b\bar{b}}}{\sigma_{jj}}$) suffice to identify the color structure of a new dijet resonance. Figs.~\ref{fig:exact_3000_and_3500} and \ref{fig:exact_4000_and_4500} show the results. Each subplot of the figures is essentially the same as the ``top view projection'' of the 3-dimensional plots in Fig. \ref{fig:3d_dcol_2_5}, but for more masses ($3.0-4.5\,\tev$) and degrees of deviations from the central values of $\dcol$ ($\dcol\pm\,20,\,50\,\%$).

In each subplot of Figs.~\ref{fig:exact_3000_and_3500} and \ref{fig:exact_4000_and_4500}, we show viable models of colorons (in blue) and leptophobic $\zp$'s (in green) leading to a set of values of $\dcol$ within $20\,\%$ (between dashed lines in the figures) and $50\,\%$ (between solid lines) given a fixed dijet cross section for each mass. The region of parameter space corresponds to the viable models predicting dijet cross sections that are not excluded by the current dijet searches and are accessible to the $14\,\tev$ LHC with $\mathcal{L}=1000\,\ifb$, with total decay widths accessible by narrow width dijet resonance searches%
\footnote{Sec. \ref{sec:paramspace} and Figs.~\ref{fig:param_space_col} and \ref{fig:param_space_zp} discuss the viable region of parameter space.}%
. In Figs.~\ref{fig:exact_3000_and_3500} and \ref{fig:exact_4000_and_4500}, the lower border (closet to the origin) of both the coloron and $\zp$ models corresponds to the limit of viable models in which the inaccessible up ratio of couplings $g_u^2/(g_u^2 + g_d^2)$ equals $0$. The upper border of the region represents the opposite limit, in which the up ratio equals its maximum value of $1$.

For illustration, we choose the common set of values of $\dcol = 0.003,\,0.007,\,0.01$ (that are allowed by the constraints) for both coloron and $\zp$ models. The sets of plots for resonances of masses $3.0\,\tev$ and $3.5\,\tev$ are shown in Fig.~\ref{fig:exact_3000_and_3500} (top and bottom panels, respectively), while similar plots for $4.0$ and $4.5\,\tev$ resonances are shown in Fig.~\ref{fig:exact_4000_and_4500}.

\begin{figure}[ht]
{
\includegraphics[width=0.85\textwidth, clip=true]{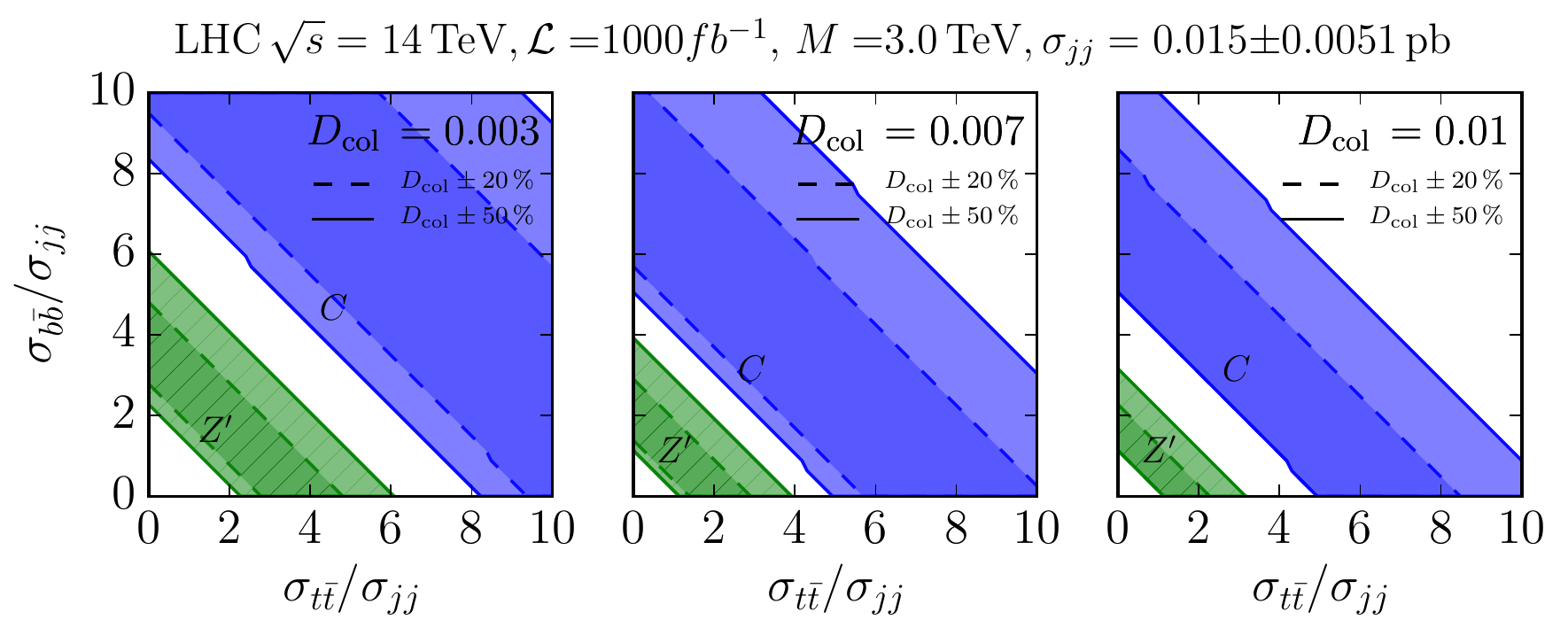}
\\
\includegraphics[width=0.85\textwidth, clip=true]{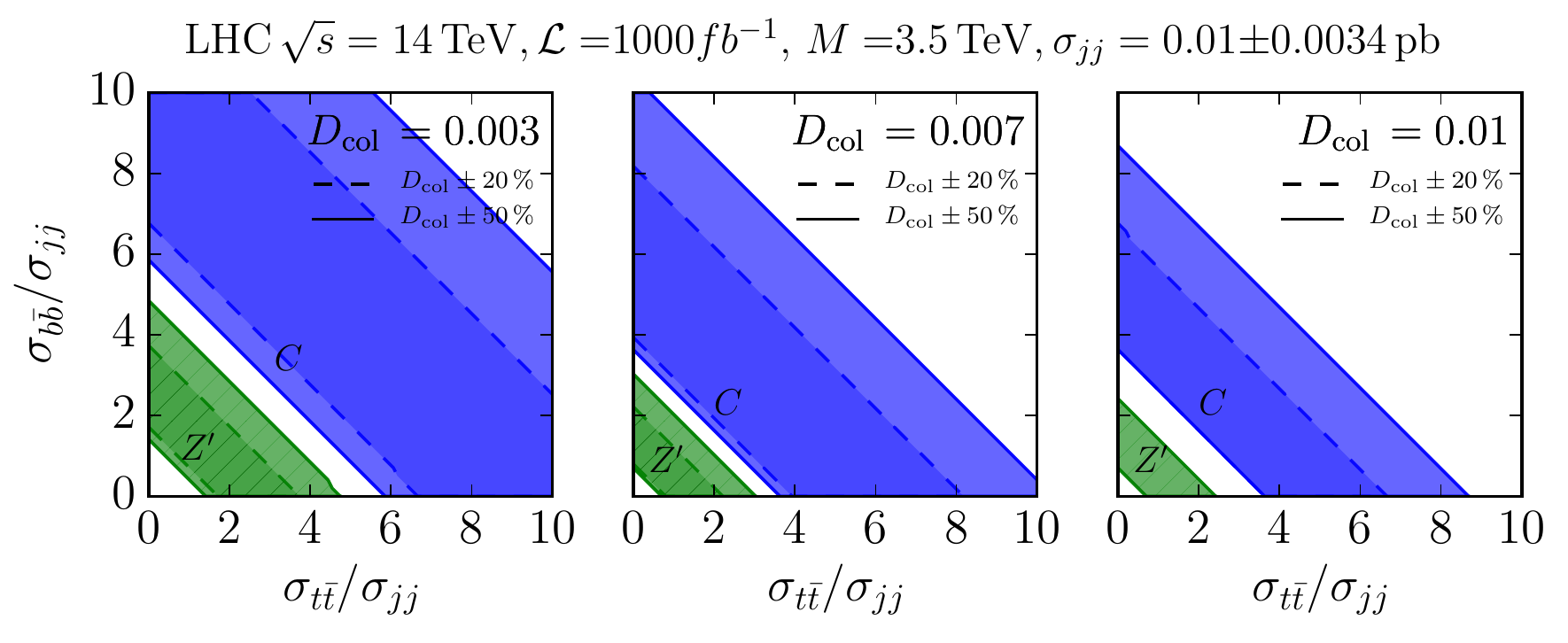}
}
\caption{%
The region of parameter space corresponding to viable models of colorons (in blue) and $\zp$ bosons (in green, with diagonal hatches) that is consistent with measurements, at the LHC with $\sqrt{s} = 14\,\tev$, of the ratios $\sigma_{t\bar{t}}$ and $\sigma_{b\bar{b}}$ for values of $\dcol$ within a $20\,\%$ (region between dashed lines) and $50\,\%$ (region between solid lines) range, for a fixed value of dijet cross section. The dijet cross section for each mass is the value allowing a $5\sigma$ discovery at the $14\tev$ LHC.
These plots illustrate the measurement precision in $\sigma_{t\bar{t}}$ and $\sigma_{b\bar{b}}$ that is required to distinguish between the coloron and the leptophobic $\zp$.  The set of plots for resonances with a mass of $3.0\,\tev$ ($3.5\,\tev$) is in the top (bottom) panel.
}
\label{fig:exact_3000_and_3500}
\end{figure}

\begin{figure}[ht]
{
\includegraphics[width=0.85\textwidth, clip=true]{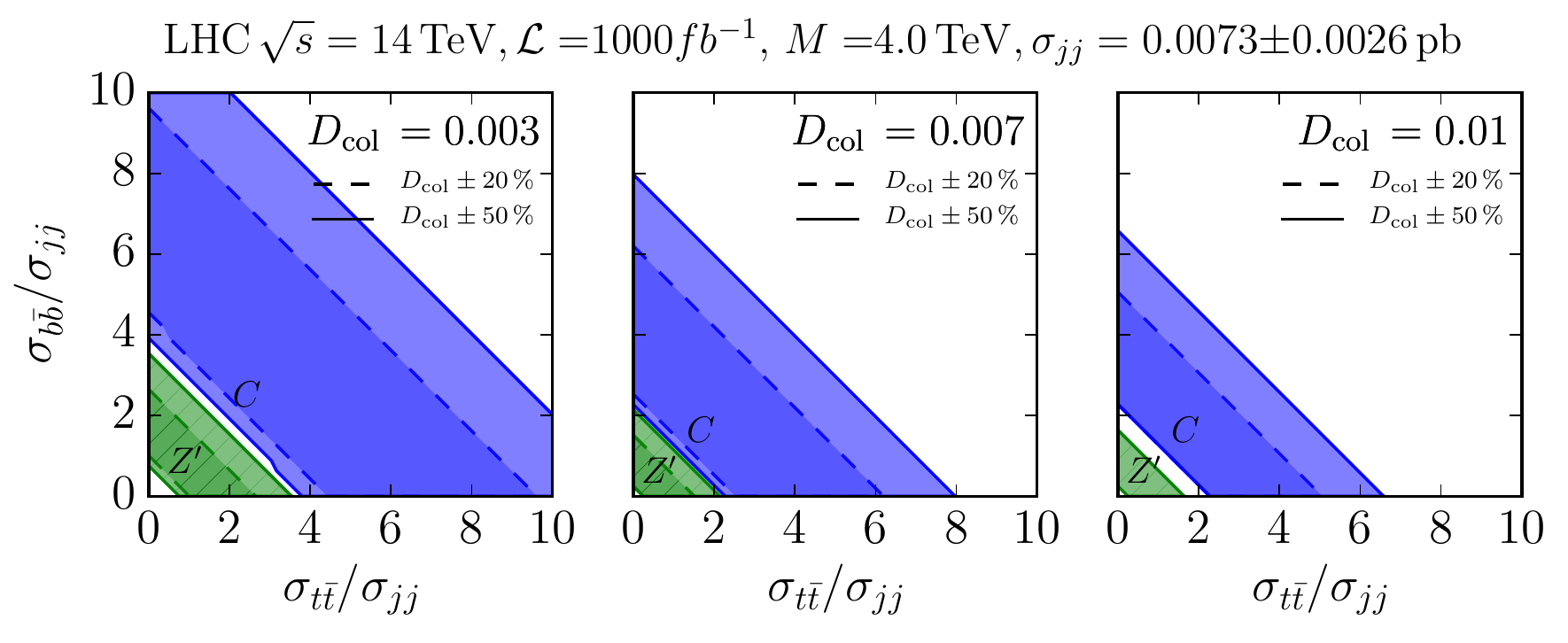}
\\
\includegraphics[width=0.85\textwidth, clip=true]{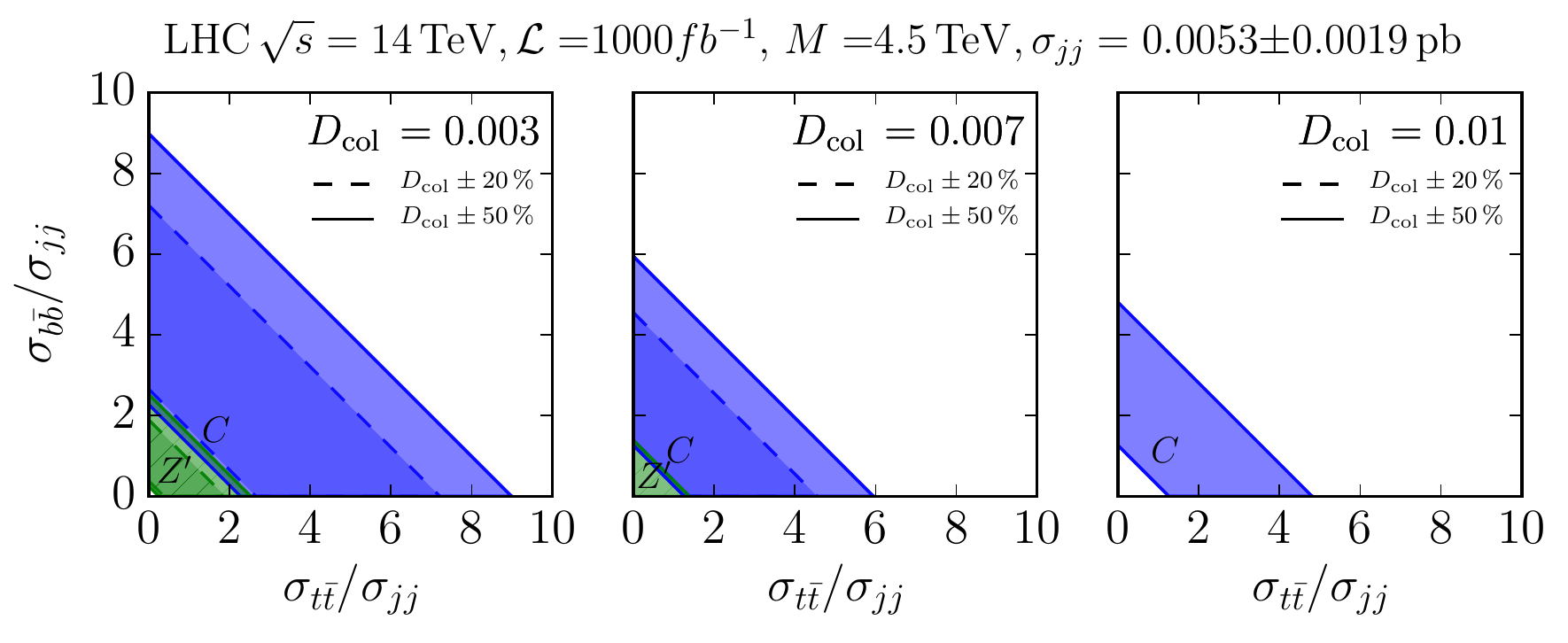}
}
\caption{Same as Fig.~\ref{fig:exact_3000_and_3500}, but for $4.0\,\tev$ (top panel) and $4.5\,\tev$ (bottom panel) resonances.
}
\label{fig:exact_4000_and_4500}
\end{figure}

\bibliography{colvszp}{}
\bibliographystyle{apsrev}

\end{document}